\pdfoutput=1
\documentclass[%
superscriptaddress,
 amsmath,amssymb,
 aps,
]{revtex4-2}

\usepackage{bm}
\usepackage{mathtools}
\usepackage{comment} 

\usepackage[dvipsnames]{xcolor}

\begin{document}


\title{Elastohydrodynamic coupling enhances\\flow generation by coordinated ciliary beating}

\author{Shota Nakano}
\affiliation{Graduate School of Engineering Science, The University of Osaka, Toyonaka 560-8531, Japan}

\author{Shinji Deguchi}
\affiliation{Graduate School of Engineering Science, The University of Osaka, Toyonaka 560-8531, Japan}

\author{Daiki Matsunaga}
\email{daiki.matsunaga.es@osaka-u.ac.jp}
\affiliation{Graduate School of Engineering Science, The University of Osaka, Toyonaka 560-8531, Japan}

\date{\today}

\begin{abstract}
Ciliary arrays pump fluid at low Reynolds number through non-reciprocal beating and phase coordination between neighbouring cilia. 
Previous studies have demonstrated that antiplectic metachronal waves are more effective than symplectic waves in enhancing transport, and have proposed several physically intuitive explanations for this preference.
What remains incomplete is a predictive analytical understanding of how hydrodynamic coupling and beat geometry determine the flow-maximising phase difference. 
Here, we address this problem in two steps: we first use reinforcement learning to identify flow-maximising coordination in a bead--spring cilia model, and then introduce an analytically tractable reduced model, termed a tilted-slider model, to analyse the weak-coupling limit.
Reinforcement learning identifies antiplectic coordination as the flow-maximising state in linear arrays, and shows that the phase difference between neighbouring cilia accounts for most of the flow enhancement.
We then use the tilted-slider model to show that a shift of the time-averaged position opposite to the effective-stroke direction enhances fluid transport through its coupling with the elastic restoring force.
The reduced model further reveals that antiplectic coordination can be optimal, consistent with previous studies, whereas symplectic coordination can instead become optimal depending on beat geometry.
These results identify a simple elastohydrodynamic mechanism underlying flow-maximising metachronal coordination.
\end{abstract}

\maketitle

\section{Introduction}\label{sec:introduction}

Motile cilia are slender, actively beating organelles that generate fluid motion in biological processes ranging from mucociliary clearance and cerebrospinal fluid transport to feeding, pumping and locomotion in microorganisms \citep{blake1974mechanics,lauga2009hydrodynamics,gilpin2020multiscale,omori2025ciliary}.
At the scale of individual ciliary beats, the surrounding fluid is typically governed by Stokes flow, where time-reversible reciprocal actuation cannot, by itself, produce net transport, as captured by the scallop theorem \citep{purcell1977life,ishimoto2012coordinate}.
A single motile cilium generates pumping by executing a non-reciprocal beat, usually consisting of an effective stroke away from the surface and a recovery stroke closer to the surface \citep{aiello1972metachronal,sanderson1981ciliary}.
In ciliary arrays, an additional level of organisation arises when neighbouring cilia beat with a phase difference, forming metachronal waves \citep{machemer1972ciliary, gheber1989synchronization, brumley2012hydrodynamic}. 
Waves propagating in the direction of the effective stroke are termed symplectic, whereas those propagating in the opposite direction are termed antiplectic \citep{knight1954relations}. 
The pumping performance of such arrays is thus characterised by a cycle-averaged flow rate that depends on both the individual beat and the phase relation between neighbouring cilia.

Autonomous or self-sustained ciliary models, such as hydrodynamically coupled oscillators \citep{cosentino2002rowers,wollin2011metachronal} or rotors, have been used to explain how phase differences can emerge between beating cilia.
This line of work identified hydrodynamic coupling as a key mechanism for ciliary synchronisation and metachronal coordination, and clarified the roles of beat geometry \citep{vilfan2006hydrodynamic} and oscillator compliance \citep{niedermayer2008synchronization}, as well as generic conditions for phase locking in coupled phase oscillators \citep{uchida2011generic}.
Extensions to chains and carpets further demonstrated travelling metachronal waves and collective phase dynamics in larger arrays \citep{uchida2010synchronization,wollin2011metachronal}.
Recent studies have clarified how global metachronal coordination is shaped by local factors, including beat and array geometry \citep{meng2021conditions,kanale2022spontaneous}, ciliary elasticity \citep{von2024hydrodynamic}, near-field hydrodynamic interactions \citep{cheng2024near}, and topological constraints on metachronal wave states \citep{chakrabarti2022multiscale}.

Motivated by the efficient transport achieved by metachronal waves in biological ciliary arrays, previous studies have asked which phase coordination maximises fluid transport when the beat kinematics are prescribed or externally controlled.
It is now well established that antiplectic coordination often generates larger fluid transport than symplectic coordination, as shown by numerical simulations \citep{gauger2009fluid,khaderi2011microfluidic,ding2014mixing,zhang2021transport,zhang2022metachronal} and experiments \citep{milana2020metachronal,dong2020}.
The same trend has also been reported in models of mucociliary transport in two-phase flows \citep{chateau2017transport,chateau2018transport,chateau2019antiplectic}, as well as in ciliated swimmers \citep{ito2019swimming,omori2020swimming}.
This antiplectic advantage is commonly interpreted as shielding or obstruction effects between neighbouring cilia.
In antiplectic coordination, a cilium undergoing the recovery stroke is exposed to the competing forward flow generated by an upstream neighbour undergoing the effective stroke, thereby reducing the backward flow \citep{gauger2009fluid,ding2014mixing,zhang2021transport,zhang2022metachronal}; this effect is often referred to as shielding of the recovery stroke.
In symplectic coordination, by contrast, a cilium in the upright position can obstruct the forward flow generated by an upstream neighbouring cilium undergoing the effective stroke \citep{khaderi2011microfluidic,milana2020metachronal}; this effect is often referred to as obstruction of the effective stroke.
Together, these mechanisms explain why antiplectic coordination can produce larger net transport than symplectic coordination.
This effective--recovery asymmetry has also been quantified in terms of the spacing between neighbouring cilia tips: antiplectic coordination produces wider tip spacing during the power stroke and narrower tip spacing during the recovery stroke, whereas symplectic coordination presents the opposite pattern \citep{dong2020}.

Overall, previous studies have provided important physical interpretations of why antiplectic coordination can enhance transport, particularly by focusing on the flow field generated during the effective and recovery strokes.
At the same time, explanations based on instantaneous configurations during the beat cycle do not by themselves predict how cycle-averaged transport changes with the phase difference, leaving a need for a theoretical framework for the flow-maximising coordination.
Such a framework might also help us understand how the flow-maximising coordination depends on system parameters, such as beat geometry and inter-cilium spacing, and may also offer a useful perspective on efficient coordination in biological ciliary arrays.
In this study, we address these problems in two steps.
First, we use reinforcement learning to identify flow-maximising coordination in a bead--spring cilia model.
Reinforcement learning has recently been used to discover and analyse actuation strategies in low-Reynolds-number systems \citep{zou2022gait,lin2024emergence,tokoro2026optimal}, and here we use it as an exploratory tool to capture the basic features of flow-maximising coordination.
Second, motivated by the resulting coordination, we introduce a reduced model, termed the tilted-slider model, to understand the underlying physical mechanism.
Inspired by rower models \citep{cosentino2002rowers,wollin2011metachronal,hamilton2021changes}, this model retains only the driving force, elastic restoring force, and hydrodynamic coupling, allowing us to examine how the flow-maximising coordination depends on inter-cilium spacing and beat geometry.
Together, these analyses lead to a physical picture in which flow enhancement arises from elastohydrodynamic coupling: hydrodynamic interactions shift the time-averaged ciliary position, and this shift modifies the elastic response over the beating cycle.

\section{Problem statement and method}
\begin{figure}
  \centering
  \includegraphics[width=0.80\linewidth]{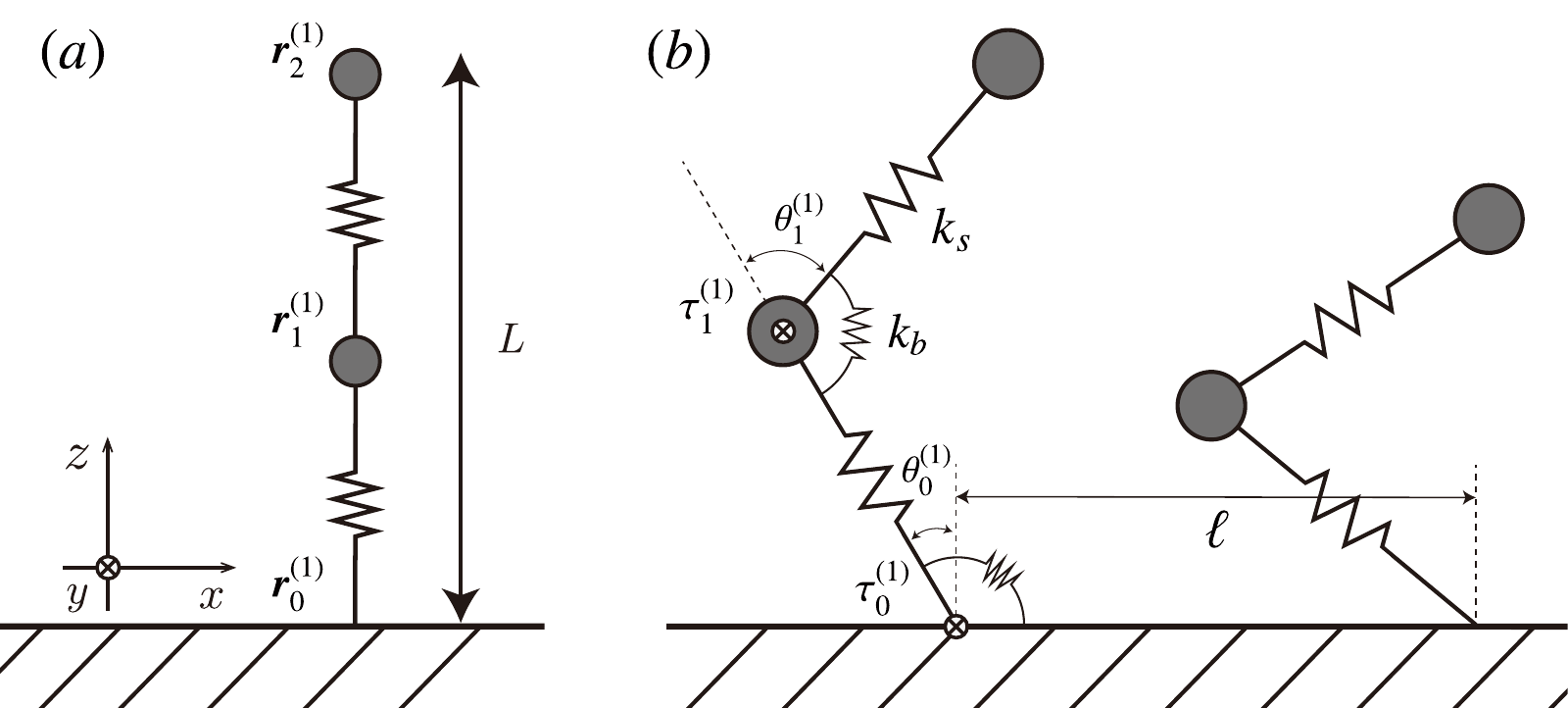}
  \caption{
    Schematics of the bead--spring based (a) cilium model ($M = 1$) and (b) two cilia model ($M = 2$). \label{fig:fig1}
  }
\end{figure}

Consider $M$ cilia models arranged in a one-dimensional array along the $x$-axis with a constant inter-cilium spacing $\ell$, and figure~\ref{fig:fig1} illustrates the problem setup for the cases (a) $M=1$ and (b) $M=2$.
Each cilium has a resting length $L$ and is modelled as a bead--spring chain anchored at its base to a planar no-slip wall at $z = 0$.
The surrounding fluid is an incompressible Newtonian fluid with the viscosity $\mu$ and the density $\rho$, and the flow is assumed to be in the Stokes regime since the Reynolds number is $Re = L^2 f \rho/\mu \ll 1$, where $f$ is the typical beating frequency.

The cilia are driven by active torques $\bm{\tau}$ applied at the joints, as illustrated in figure~\ref{fig:fig1}(b).
The main objective here is to understand the collective motion of cilia that maximises the net fluid pumping in $x$-direction; in other words, to elucidate the time-dependent active torque inputs $\bm{\tau} (t)$ that maximise the pumping.
Note that although we solve the full three-dimensional dynamics of the system, all beads and springs remain in the same $xz$-plane due to symmetry; the active torques are applied only about the $y$-axis.
Reinforcement learning framework is introduced to search for a flow-maximising motion.

\subsection{Cilia model}

The minimal cilium model in this work consists of two spherical beads of radius $a$, connected by extensional and bending springs.
Related discrete bead models of cilia have been used in previous studies  \citep[e.g.][]{kim2006pumping, osterman2011finding, elgeti2013emergence}.
As shown in figure~\ref{fig:fig1}(a), the shape of the $m$-th cilium is represented by three position vectors $\bm{r}_i^{(m)}$, 
where the superscript $m$ denotes the cilium index and the subscript $i$ denotes the local bead index within each cilium: $\bm{r}_0^{(m)}$ is the basal position, $\bm{r}_1^{(m)}$ is the position of the middle bead, and $\bm{r}_2^{(m)}$ is the position of the tip bead.
The root angle $\theta_0^{(m)}$ and the mid-joint angle $\theta_1^{(m)}$ are defined as signed angles in $(-\pi,\pi)$ by $\cos\theta_0^{(m)}=\hat{\bm r}_{01}^{(m)}\cdot\bm e_z$ and $\sin\theta_0^{(m)}=\bm e_y\cdot(\bm e_z\times\hat{\bm r}_{01}^{(m)})$;
$\cos\theta_1^{(m)}=\hat{\bm r}_{01}^{(m)}\cdot\hat{\bm r}_{12}^{(m)}$ and $\sin\theta_1^{(m)}=\bm e_y\cdot(\hat{\bm r}_{01}^{(m)}\times\hat{\bm r}_{12}^{(m)})$,
where $\hat{\bm r}_{ij}^{(m)}=(\bm r_j^{(m)}-\bm r_i^{(m)})/\|\bm r_j^{(m)}-\bm r_i^{(m)}\|$ is the unit vector along the segment connecting $i$- and $j$-th beads, 
and $\bm{e}_x$, $\bm{e}_y$, $\bm{e}_z$ are the unit vectors of the Cartesian coordinates.

The force $\bm{F}_i^{(m)}$ acting on $i$-th bead of the $m$-th cilium consists of passive/elastic contributions from the extensional and bending springs, as well as active contributions generated by the joint torques $\bm{\tau}$.
The potential energy of the extensional springs is defined as
\begin{equation}
U_s^{(m)} = \frac{1}{2}k_s \sum_{i=1}^2 \left( \|\bm{r}_i^{(m)} - \bm{r}_{i-1}^{(m)}\| - \frac{L}{2}\right)^2.
\label{eq:Us}
\end{equation}
where $k_s$ is the extensional stiffness and $L/2$ is the rest length of each segment.
The bending potential energy is defined as
\begin{equation}
U_b^{(m)} = \frac{1}{2}k_b \left(\alpha\,(\theta_0^{(m)})^2 + (\theta_1^{(m)})^2\right),
\label{eq:Ub}
\end{equation}
where $\alpha = 2$ is a prefactor introduced to match the relaxation timescales of the basal and middle joints, as derived in Appendix~\ref{app:alpha}.
The active torques $\bm{\tau}^{(m)} = (\bm{\tau}_0^{(m)}, \bm{\tau}_1^{(m)})$ are applied at the base $\bm{\tau}_0^{(m)} = \tau_0^{(m)} \bm{e}_y$ and the middle bead $\bm{\tau}_1^{(m)} = \tau_1^{(m)} \bm{e}_y$. 
The torque magnitudes are bounded by $-\tau_{\max} \leq \tau_k^{(m)} \leq \tau_{\max}$ where $\tau_{\max}$ is the maximum torque amplitude.
We now introduce an effective potential
\begin{equation}
U_a^{(m)} = - \left(\alpha\,\tau_0^{(m)} \theta_0^{(m)} + \tau_1^{(m)} \theta_1^{(m)}\right),
\label{eq:Wa}
\end{equation}
which satisfies $-\partial U_a^{(m)}/\partial \theta_i^{(m)} = \tau_i^{(m)}$, so that the force can be written in the compact form.
The total force acting on the $i$-th bead ($i=1, 2$) is given by
\begin{equation}
\bm{F}_i^{(m)} = - \frac{\partial }{\partial \bm{r}_i^{(m)}}\left(U_s^{(m)} + U_b^{(m)} + U_a^{(m)}\right).
\label{eq:Fi_total}
\end{equation}

\subsection{Fluid mechanics}\label{subsec:fluid}
We consider an incompressible Newtonian fluid in the low-Reynolds-number regime, for which the fluid velocity field $\bm{u}$ and pressure $p$ satisfy the Stokes equations,
\begin{equation}
-\nabla p + \mu \nabla^{2}\bm{u}=\bm{0},
\qquad 
\nabla\cdot\bm{u}=0,
\label{eq:stokes}
\end{equation}
and the flow is assumed to be quiescent at infinity.
When an isolated point force $\bm{F}$ acts at $\bm{y}$ under the free-space, the resulting velocity field at $\bm{x}$ is 
$\bm{u}(\bm{x})=\bm{J}(\bm{x},\bm{y})\cdot\bm{F}(\bm{y})$, where $\bm{J}(\bm{x},\bm{y})$ is the free-space Green's function for a point force.
In the presence of an infinite planar no-slip wall at $z=0$, the Green's function is given by
$\bm{G}(\bm{x},\bm{y})=\bm{J}(\bm{x},\bm{y})+\bm{J}^w(\bm{x},\bm{y})$,
where $\bm{J}^w(\bm{x},\bm{y})$ denotes the corresponding image system \citep{blake1971note}.
To model the bead dynamics, we employ a simplified form of Stokesian dynamics in which only translational motion is retained and rotational degrees of freedom are neglected.
The mobility tensor of hydrodynamically interacting spheres close to a planar wall can be derived using the Fax\'en relationship \citep{swan2007simulation}, and the translational velocity of $i$-th bead at position $\bm{x}$ is given as
\begin{equation}
  \begin{split}
    \dot{\bm{r}}_i (\bm{x}) =& \left[\dfrac{\bm{I}}{6 \pi \mu a} + \left(1+\dfrac{a^2}{6} \nabla^2_{\bm{x}} \right)\left(1+\dfrac{a^2}{6} \nabla^2_{\bm{y}} \right)\bm{J}^w(\bm{x},\bm{y})\bigg|_{\bm{x}=\bm{y}=\bm{r}_i}  \right] \cdot \bm{F}_i (\bm{y})\\
    & +\sum_{j\neq i} \left(1+\dfrac{a^2}{6} \nabla^2_{\bm{x}} \right)\left(1+\dfrac{a^2}{6} \nabla^2_{\bm{y}} \right)\bm{G}(\bm{x},\bm{y})\bigg|_{\bm{x}=\bm{r}_i,\bm{y}=\bm{r}_j} \cdot \bm{F}_j (\bm{y}),
  \end{split}
  \label{eq:bead_velocity}
\end{equation}
where $\bm{I}$ is the identity tensor and $\bm{F}_j$ denotes the force acting on $j$-th bead at position $\bm{y}$.
Note that, only in Eq.~\eqref{eq:bead_velocity}, the indices $i$ and $j$ denote global bead indices running from $1$ to $2M$.
The operators $\nabla^2_{\bm{x}}$ and $\nabla^2_{\bm{y}}$ denote Laplacian with respect to the positions $\bm{x}$ and $\bm{y}$, respectively.

The instantaneous volume flow rate in the \(+x\) direction is given by
\begin{equation}
  q(t) = \sum_{m=1}^M \sum_{i=1}^2 \frac{z_i^{(m)}(t)\,F_{x,i}^{(m)}(t)}{\pi \mu},
  \label{eq:qx_reward}
\end{equation}
where $F_{x,i}(t)$ is the $x$-component of the force acting on $i-$th bead located at height $z_i(t)$ above the wall \citep{liron1978fluid,smith2008modelling}.
Given the bead forces and trajectories, we quantify the net fluid transport by the time-averaged flow rate over a single period $T$ as
\begin{equation}
  Q_M = \dfrac{1}{T} \int_0^T q(t)\, \mathrm{d}t.
\end{equation}

\subsection{Nondimensionalization and numerical methods}
The governing equations are nondimensionalized using the cilium length $L$ as the length scale, the torque bound $\tau_{\max}$ as the torque scale, and the dynamic viscosity $\mu$ as the viscosity scale.
In this work, three dimensionless parameters govern the dynamics: the dimensionless bead radius $a^* = a/L$, the dimensionless extensional stiffness $k_s^* = k_s L^2/\tau_{\max}$, and the dimensionless bending stiffness $k_b^* = k_b/\tau_{\max}$.
The first parameter controls the relative strength of the hydrodynamic interactions between beads, whereas the second and third parameters characterise the stretching and bending stiffnesses relative to the maximum active torque, respectively.

Two dimensionless parameters are fixed throughout this work, $k_s^* = 1.0\times10^{4}$ and $a^* = 0.05$, while the dimensionless bending stiffness is varied over $k_b^* \in [0.8, 1.2]$.
A large value of $k_s^*$ is used to enforce near-inextensibility, so that the inter-bead distances remain approximately constant throughout the simulations.
The bead positions are updated by integrating equation~\eqref{eq:bead_velocity} in time using the standard forward Euler scheme with a time step $\Delta t^* = 4.0\times10^{-5}$.
As also explained in the next subsection, the active torque $\bm{\tau}$ is updated by reinforcement learning algorithm at every time interval $T_a^* = 2.0\times10^{-2}$, which we refer to as the action interval \citep{tokoro2026optimal}.

\subsubsection{Reinforcement learning}
In this work, we introduced a reinforcement learning algorithm, proximal policy optimisation (PPO) \citep{schulman2017PPO}, to search for a strategy to maximise the pumping.
Reinforcement learning aims to learn a policy that selects an action $A_t$ based on the current state $S_t$ to maximise the expected cumulative reward $R_{\mathrm{ep}}$, defined by the immediate reward $R_t$ \citep{sutton1998reinforcement}.
In this work, the active torques $\tau_k^{(m)*} = \tau_k^{(m)}/\tau_\mathrm{max} \in [-1, 1]$ are set as the action $A_t = (\tau_0^{(1)*}, \tau_1^{(1)*}, \cdots, \tau_0^{(M)*}, \tau_1^{(M)*})$, the instantaneous cilia configuration as the current state $S_t$ as explained later, and the dimensionless instantaneous flow rate as the immediate reward $R_t = q^*(t) = q (t) \mu / \tau_\mathrm{max}$.
The action $A_t$ is updated at intervals of $T_a^*$; in other words, the action is held constant for $T_a^*/\Delta t^*$ time steps.
Through the process of maximising the expected cumulative reward
\begin{equation}
  R_{\mathrm{ep}} = \sum_{k=t}^{N-1} \gamma^{k-t} R_k,
\end{equation}
the PPO algorithm searches for the input torques that maximise the net pumping in the $+x$ direction, over an episode of length $N$ steps.
Note that we utilised the discount factor $\gamma=0.997$ and the total step $N = 10^4$, and the episode length $t^* = N T_a^* = 2.0 \times 10^2$ is sufficiently large compared with the beating period.

To ensure training stability, different state representations $S_t$ are introduced for the cases $M = 1$ and $M > 1$.
For a single cilium $M=1$, the state $S_t = s_t^{(1)}$ is used, where $s_t^{(m)}$ denotes the set of variables describing the instantaneous kinematics of the $m$-th cilium as
\begin{equation}
s_t^{(m)}=\bigl(\cos\theta_0^{(m)},\sin\theta_0^{(m)},\dot\theta_0^{(m)},\cos\theta_1^{(m)},\sin\theta_1^{(m)},\dot\theta_1^{(m)}\bigr),
\end{equation}
and $\dot{\theta}_k^{(m)}$ is estimated using a finite difference between successive intervals $T_a^*$.
For $M>1$, the state is defined as $S_t=\bigl(s_t^{(1)},d^{(1)},s_t^{(2)},\cdots,d^{(M-1)},s_t^{(M)}\bigr)$ where variable $d^{(m)}=(\cos (\theta_0^{(m+1)}-\theta_0^{(m)}), \sin(\theta_0^{(m+1)}-\theta_0^{(m)})$) represents the angle difference with the neighbouring cilia.

After the optimisation has converged, the optimised motion is obtained from an evaluation rollout.
The beating periods $T^*$ and the inter-ciliary phase difference $\Delta$ are evaluated from the Fourier transform of the $x$-coordinate of $\bm{r}_2^{(m)}$; $T^*$ is obtained as the inverse of the frequency corresponding to the largest power spectral component, while $\Delta$ is defined as the phase difference between cilia at the dominant frequency, as explained in detail below.

\section{Exploring flow-maximising coordination using reinforcement learning} \label{sec:cilia}

In this section, we present our first main result, in which reinforcement learning is used to explore optimal collective beating of cilia arrays.
We begin with a single cilium system ($M=1$), and then examine coordinated beating in cilia arrays ($M>1$).
We further focus on the two cilia system ($M=2$) for a more detailed examination of the parameter dependence and beating patterns.
For each parameter set, the RL optimisation was repeated in five independent training runs with different random seeds.
Unless otherwise noted, reported values are means over these five runs, and error bars indicate the standard deviation across runs.

\begin{figure}
  \centering
  \includegraphics[width=0.8\linewidth]{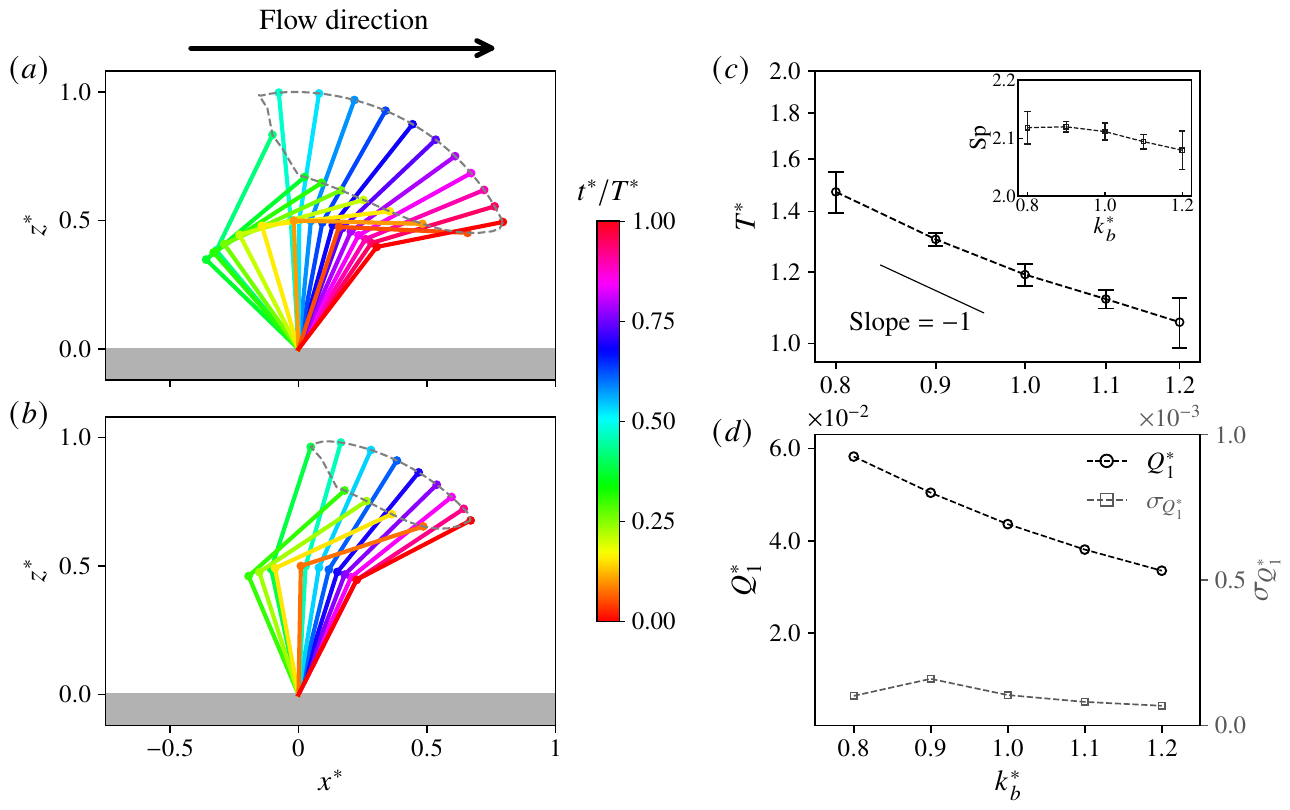}
\caption{
Beating of a single cilium model optimised by reinforcement learning.
(a,b) Representative optimised beats over one period with different bending stiffness (a) $k_b^*=0.8$ and (b) $k_b^*=1.2$.
Colours indicate the phase $t^*/T^*$, and the grey strip denotes the no-slip wall at $z^*=0$.
(c) Dimensionless beating period $T^*$ of a cilium with different $k_b^*$.
The inset figure shows the corresponding effective sperm number $\mathrm{Sp}$ evaluated using \eqref{eq:Sp_eff_closed}.
(d) Dimensionless flow rate $Q_1^*$ for different $k_b^*$ (left axis); 
the corresponding standard deviation across the five independent training runs, denoted by $\sigma_{Q^*_1}$, is shown on the right axis. \label{fig:fig2}}
\end{figure}

\subsection{Beating of a single cilium}\label{sec:single_cilium}
We first show the optimised strokes of a cilium $M=1$ for different bending stiffnesses $k_b^*$ in figures~\ref{fig:fig2}(a) and (b), and supplemental movie 1.
The cilium exhibits a non-reciprocal beating motion consisting of effective and recovery strokes, reminiscent of the motion of biological motile cilia \citep[e.g][]{blake1974mechanics}. 
It is noteworthy that the optimised beating pattern is similar to that of biological cilia, despite the absence of prior knowledge of natural ciliary beating.
The cilium with the larger bending stiffness $k_b^*$ exhibits a smaller beating trajectory, as expected.
Figure~\ref{fig:fig2}(c) shows the beating period $T^*$ is inversely proportional to the bending stiffness $k_b^*$, and a stiffer cilium beats faster.
Figure~\ref{fig:fig2}(d) shows the cycle-averaged volume flow rate $Q_1^*$ together with its standard deviation $\sigma_{Q^*_1}$ across training runs.
The value $\sigma_{Q_1^*}$ is plotted on the right-hand vertical axis, and its magnitude is small compared with the mean flow rate, with $\sigma_{Q_1^*}/Q_1^* \sim O(10^{-3})$, whereas the relative variation in the beating period is $O(10^{-2})$. 
This result indicates that each training converges to slightly different beating motions that nevertheless produce nearly the same flow rate \(Q_1^*\).

The emergence of the optimal beating period $T^*$ can be understood as a consequence of the competition between the elastic relaxation timescale and the actuation timescale \citep{gauger2009fluid,eloy2012kinematics,tokoro2026optimal}.
This competition can be characterised with the sperm number \citep{lagomarsino2003simulation}, which is the dimensionless parameter defined as
\begin{equation}
  \mathrm{Sp} \coloneqq L\left(\frac{\omega\,\xi_{\perp}}{B}\right)^{1/4},
  \label{eq:Sp_def}
\end{equation}
where $L$ is the cilium length, $\omega=2\pi/T$ is the angular beating frequency, $\xi_{\perp}$ is the normal drag coefficient per unit length, and $B$ is the bending rigidity.
At small sperm numbers, the flow rate is small because the beating period is long compared with the relaxation time of the cilium-fluid system, resulting in a limited number of effective strokes per unit time.
At large sperm numbers, the flow rate is also small because the beating period is short relative to the relaxation time, leaving insufficient time for the deformation to fully develop before the actuation reverses.
An optimal flow rate therefore arises at an intermediate sperm number, where these two timescales are comparable.
For the present discrete two-link bead--spring model, we approximate the value $B$ as $B \approx k_bL/2$ from the discrete bending energy, and the value $\xi_{\perp}$ as $\xi_{\perp} \approx 12\pi\mu a/L$ from a local-drag estimate based on the two beads, and these give
\begin{equation}
  \mathrm{Sp} \approx
  \left(\frac{48\pi^{2}\mu a\,L^{2}}{k_b\,T}\right)^{1/4}.
  \label{eq:Sp_eff_closed}
\end{equation}
The inset of figure~\ref{fig:fig2}(c) shows that the sperm number is nearly constant $\mathrm{Sp} \sim 2$, over this parameter range; this also explains why the period $T^*$ scales inversely with the stiffness $k_b^*$, as shown in Fig.~\ref{fig:fig2}(c).

\begin{figure}
  \centering
  \includegraphics[width=0.8\linewidth]{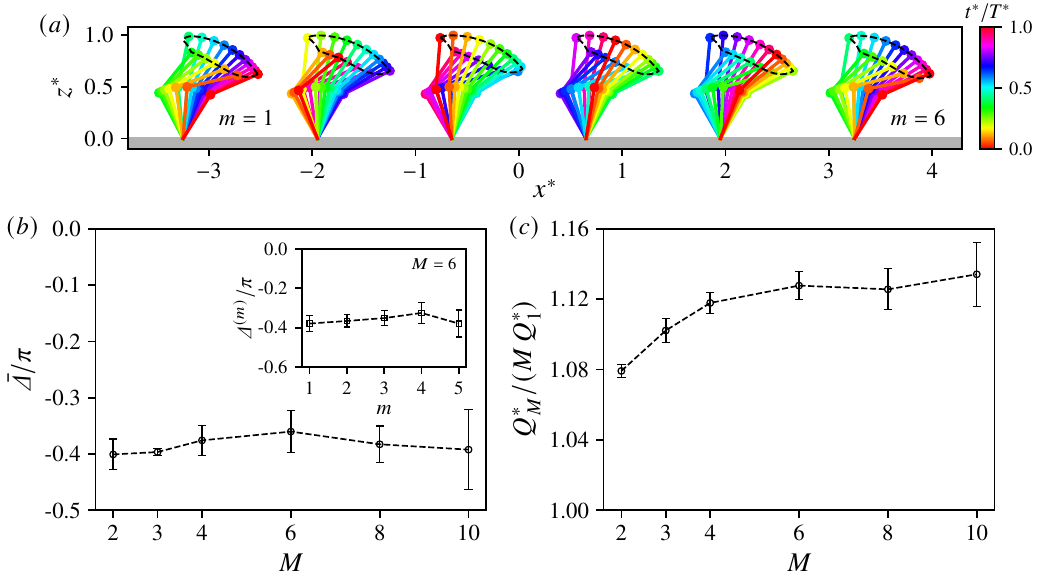}
\caption{
Flow-maximising coordination of cilia arrays optimised by reinforcement learning, under $k_b^*=1.2$ and $\ell^*=1.3$.
(a) Representative beating pattern for $M=6$ over one period.
Colours indicate the phase within the cycle, $t^*/T^*$, and the grey strip denotes the no-slip wall.
(b) Array-averaged nearest-neighbour phase difference, $\bar{\Delta}/\pi$ for different numbers of cilia $M$.
Inset: pairwise phase difference $\Delta^{(m)}/\pi$ for the $M=6$ array.
(c) Flow enhancement ratio $Q_M^*/(M Q_1^*)$ for different $M$,
where $Q_M^*$ and $Q_1^*$ denote the flow rates of the optimised $M$ cilia and single-cilium systems, respectively.
Symbols and error bars in (b,c) show the mean and standard deviation over five independent training runs.
\label{fig:fig3}
}
\end{figure}

\subsection{Coordination in linear ciliary arrays}
Next, we analyse the optimal collective motion of cilia arrays by varying number of cilia $M$, while fixing other conditions $k_b^*=1.2$ and $\ell^* = 1.3$.
As shown as a representative example with $M=6$ in figure~\ref{fig:fig3}(a) and supplemental movie 2, the individual cilia retain the effective--recovery asymmetry found in the single cilium, while adjacent cilia exhibit finite phase differences and form the metachronal coordination.
The phase difference of the beating $\Delta^{(m)}$ between the $m$- and $(m+1)$-th cilia is now evaluated as
\begin{equation}
\Delta^{(m)}=\psi^{(m)}-\psi^{(m+1)}, \qquad m=1,\ldots,M-1,
\end{equation}
where $\psi^{(m)}$ is the phase of the dominant Fourier mode of the $m$-th cilium; with this convention, $\Delta^{(m)}>0$ corresponds to symplectic coordination while $\Delta^{(m)}<0$ corresponds to antiplectic coordination.
As shown in the inset of figure~\ref{fig:fig3}(b), the phase differences for the $M=6$ array displayed in figure~\ref{fig:fig3}(a) are negative $\Delta^{(m)} < 0$ and approximately uniform without noticeable end effects.
Figure~\ref{fig:fig3}(b) shows the averaged phase difference among the cilia array $\bar{\Delta} = 1/(M-1) \sum_m^{M-1} \Delta^{(m)}$, and the value $\bar{\Delta} \simeq -0.4 \pi < 0$ remains negative and varies only weakly with the cilia number $M$.
This result suggests that reinforcement learning likewise selects antiplectic coordination, consistent with many previous studies reviewed in \S\ref{sec:introduction}, and the phase difference has weak dependence on the cilia number $M$ in agreement with \citet{gauger2009fluid}.
We next quantify how the normalized per-cilium transport $Q_M^*/(M Q_1^*)$ varies with cilia number in figure~\ref{fig:fig3}(c), where $Q_1^*$ is the flow rate of the single cilium obtained in the previous subsection; the quantity $Q_M^*/(M Q_1^*)$ can also be interpreted as the relative flow enhancement due to the coordination.
The figure shows that there is already flow enhancement of 8\% even for two cilia, and this ratio increases with the cilia number but saturates at approximately 12\% for $M \approx 6$.
\citet{gauger2009fluid} likewise reported that the flow enhancement ratio saturates, reaching approximately 40\% for $M=8$--$16$.
Note that the difference in the enhancement ratio is attributable to the difference in the cilia model, the beating pattern, and the inter-cilium spacing. 

In summary, we understand by varying the number of cilia that the basic character of the coordination can already be captured by the two cilia system. 
Although the coordination would be slightly altered by the number of cilia, the long-range hydrodynamic interactions are screened by the wall effect \citep{blake1971note,uchida2011generic}, and the optimal coordination is determined primarily by local interactions between neighbouring cilia.
We hereafter focus on the two cilia system ($M=2$) as the minimal unit for understanding coordination.

\begin{figure}
  \centering
  \includegraphics[width=0.8\linewidth]{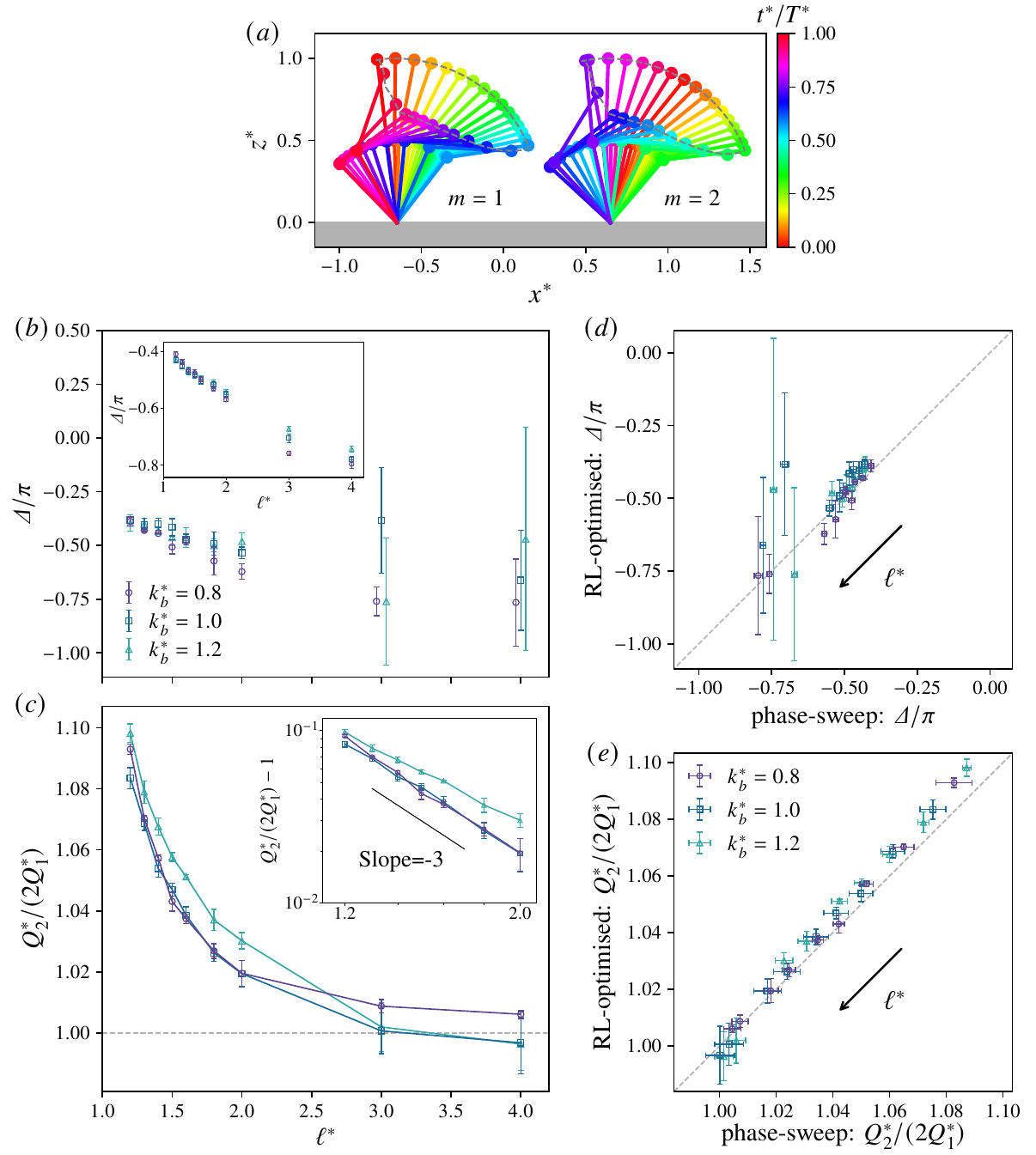}
\caption{
Beating of two cilia optimised by reinforcement learning.
(a) Representative beating patterns for $k_b^*=0.8$ and $\ell^*=1.3$.
(b) Optimal phase difference $\Delta$ as a function of inter-cilium spacing $\ell^*$.
Inset: the same plot obtained using the phase-sweep analysis.
(c) Flow enhancement ratio $Q_2^*/(2Q_1^*)$ as a function of $\ell^*$.
Inset: log--log plot of the excess flow ratio $Q_2^*/(2Q_1^*)-1$.
(d,e) Comparison between RL and the phase-sweep analysis for (d) the flow-rate ratio $Q_2^*/(2Q_1^*)$ and (e) the optimal imposed offset $\Delta/\pi$.
In (d,e), the vertical error bars indicate the standard deviation over five runs, whereas the horizontal error bars indicate the standard deviation over phase-sweeps using five different torque sequences optimised for a single cilium. \label{fig:fig4}
}
\end{figure}

\subsection{Parameter dependent coordination of two cilia} \label{subsec:twocilia}
We focus on the collective beating of two cilia $M=2$, as shown as the representative example in figure~\ref{fig:fig4}(a), and examine how the collective enhancement depends on the inter-cilium spacing $\ell^*$ and the bending stiffness $k_b^*$.
Note that the phase difference is hereafter denoted simply by $\Delta = \Delta^{(1)}$ for the two cilia setup.
Figure~\ref{fig:fig4}(b) shows that the antiplectic coordination $\Delta < 0$ is preferred regardless of $\ell^*$ and $k_b^*$, once again consistent with the previous reports reviewed in \S\ref{sec:introduction}. 
The phase difference is approximately $-0.4 \pi$ when the inter-cilium spacing $\ell^*$ is small, and this difference grows as $\ell^*$ increases.
Note that the error bars of five runs become more pronounced for $\ell^* \gtrsim 3$, because the differences in instantaneous flow $q^*$ are so small in this regime that it becomes difficult for reinforcement learning system to identify a unique optimal phase relation.
Nevertheless, the mean selected phase difference remains on the antiplectic side, consistent with the weak but persistent antiplectic preference at larger cilium spacings \citep{khaderi2011microfluidic}.
Figure~\ref{fig:fig4}(c) shows that the flow enhancement ratio $Q_2^*/(2 Q_1^*)$ is approximately 10\% when the spacing is small and decreases as the spacing increases; for large spacing as $\ell^* \gtrsim 3$, the generated flow is well approximated as being proportional to the number of cilia $M$.
The excess flow ratio $Q_2^*/(2Q_1^*) - 1$ decays rapidly with $\sim \ell^{-3}$ as shown in the inset. 
It is interesting to note that both $\Delta$ and $Q_2^*/(2Q_1^*)$ depend only weakly on the bending stiffness $k_b^*$.

The results for two cilia naturally raise a more specific question: does the flow enhancement due to the coordination require independent optimisation for each cilium, or can it already be realised by identical beating patterns with an appropriate phase difference?
To clarify this point, we perform a simple simulation in which the torque sequence optimised for a single cilium ($M=1$) is copied to the two cilia system ($M=2$), with a prescribed phase difference between the two cilia.
Let $\bm{\tau}_{\mathrm{s}}(t)=\bigl(\tau_{0,\mathrm{s}}(t),\tau_{1,\mathrm{s}}(t)\bigr)$ denote the sequence learned for the single cilium system.
We then drive the two cilia system with
\begin{equation}
\bm{\tau}^{(1)}(t;\delta)=\bm{\tau}_{\mathrm{s}}\!\left(t+\delta T_{\mathrm{s}}/2\pi\right),\qquad
\bm{\tau}^{(2)}(t)=\bm{\tau}_{\mathrm{s}}(t),
\label{eq:phase_sweep_drive}
\end{equation}
where $T_{\mathrm{s}}$ is the period of a single cilium and $\delta$ is the manually imposed phase difference.
For each $(\ell^*,k_b^*)$ condition, we swept the imposed phase difference $\delta$ to obtain the phase that maximises the flow: i.e. $\Delta = \arg\max_{\delta \in [-\pi, \pi]} Q_2^* (\delta)$.
Hereafter, we refer to this approach as the phase-sweep analysis.
Figures~\ref{fig:fig4}(d) and (e) compare the two quantities, $\Delta$ and $Q_2^*/(2Q_1^*)$, obtained by two approaches: the phase-sweep analysis on the horizontal axis and reinforcement learning on the vertical axis.
The phase-sweep analysis yields nearly the same optimal phase difference $\Delta$ and flow enhancement ratio $Q_2^*/(2Q_1^*)$ as reinforcement learning, suggesting that the flow enhancement can be explained primarily by the phase difference rather than by the independent optimisation for each cilium.
Although reinforcement learning gives slightly larger $Q_2^*/(2Q_1^*)$, this difference remains small, indicating that independent optimisation for each cilium has only a small effect.
The inset of figure~\ref{fig:fig4}(b) shows that the magnitude of the optimal phase difference $|\Delta|$ grows monotonically with the inter-cilium spacing $\ell^*$ and approaching $\Delta = -\pi$, a trend that was also suggested by reinforcement learning but was less evident because of the large uncertainty for $\ell^* \gtrsim 3$.

\section{Mechanism of flow enhancement}
In the previous section, reinforcement learning showed that antiplectic coordination is preferred over the explored parameter range of cilia number, cilia spacing, and bending stiffness.
Although reinforcement learning can search for the flow-maximising coordination under each condition, it does not by itself reveal the underlying physics of the obtained solution.
This makes it difficult to assess the generality of the obtained coordination and, hence, to understand how the optimal coordination varies across parameter space, including the transition between symplectic and antiplectic coordination.

To address this limitation, we now introduce a reduced tilted-slider model in which each cilium is represented by a bead constrained to move along a tilted straight line near a no-slip wall.
This model is inspired by rower-type minimal models for hydrodynamically coupled cilia, 
which have proved useful for describing synchronisation and metachronal coordination \citep{cosentino2002rowers,wollin2011metachronal,hamilton2021changes}.
In contrast to these rower-type models, which employ geometric switching between stroke states, the present model is driven by prescribed phase-lagged forcing together with an elastic restoring response.
The tilt angle serves as a reduced descriptor of the mean beating orientation relevant to wall-bounded pumping, while the detailed effective--recovery stroke asymmetry is ignored.
In this section, we use this reduced model to reveal analytically how the flow-maximising coordination depends on the parameters.

\begin{figure}
    \centering
    \includegraphics[width=0.6\linewidth]{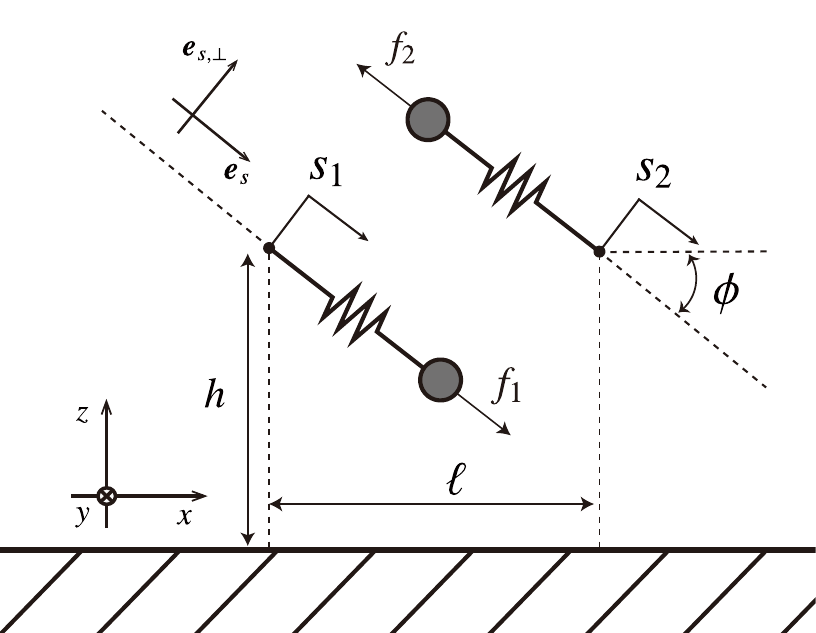}
    \caption{
        Schematic of the reduced tilted-slider model.
        Two beads of radius $a$ move along straight slider axes in the $xz$-plane at height $h$ above a no-slip wall.
        The axes are separated by a distance $\ell$ and are inclined by an angle $\phi$.
        The coordinates for each slider are denoted by $s_1$ and $s_2$, and the tangential arrows indicate the active driving forces.}
    \label{fig:fig5}
\end{figure}

\subsection{Reduced tilted-slider model}\label{subsec:tilted-slider-model}
The reduced model considered here consists of two spherical beads of radius $a$, separated by a distance $\ell$ along the $x$-direction.
Each bead is constrained to move along a straight slider axis in the $xz$-plane, and the slider axis is obtained by rotating $\bm{e}_x$ by an angle $\phi \in (-\pi/2, \pi/2)$ about the positive $y$-axis, as shown in figure~\ref{fig:fig5}.
The unit vector along the slider axis is $\bm e_s=(\cos\phi,0,-\sin\phi)$ and the perpendicular unit vector is $\bm e_{s,\perp}=(\sin\phi,0,\cos\phi)$.
The position of $i$-th bead is written as $\bm r_i=\bm c_i+s_i\bm e_s$ $(i=1,2)$, where $s_i(t)$ is the scalar displacement along the slider axis, and $\bm c_1=-\ell \bm e_x/2+h\bm e_z$ and $\bm c_2=\ell \bm e_x/2+h\bm e_z$ denote the centres of the two slider axes, located at a distance $h$ away from the wall.
The system is driven by the active forces defined as
\begin{equation}
f_1(t)=f_A\cos(\omega t+\delta),\qquad
f_2(t)=f_A\cos(\omega t),
\end{equation}
where $f_A$ is the forcing amplitude, $\omega$ is the driving angular frequency, and $\delta$ is the imposed phase difference.
Each bead is attached to its reference position by a linear restoring spring of stiffness $k_r$.
The total force on the $i$-th bead is therefore written as
\begin{equation}
\bm F_i=\bigl(f_i-k_r s_i\bigr)\bm e_s+\Lambda_i\bm e_{s,\perp},
\qquad i=1,2,
\label{eq:slider_force}
\end{equation}
where the second term represents the constraint force $\Lambda_i$ that keeps the bead on the prescribed slider axis.
The bead velocities are governed by the wall-corrected mobility relation.
The governing equations are nondimensionalized using $h$, $f_A$, and $\mu h^2/f_A$ as the characteristic length, force, and time scales, respectively.
Accordingly, time and frequency are written in dimensionless form as $t^*=f_A t/(\mu h^2)$ and $\omega^*=\mu h^2\omega/f_A$.
The reduced model is then characterised by three dimensionless parameters: the bead radius $a^*=a/h$, the inter-slider spacing $\ell^*=\ell/h$, and the spring stiffness $k_r^*=k_r h/f_A$. 
The first two characterise the geometry relative to the reference height, while the third measures the spring stiffness relative to the forcing amplitude.
Note that, as expected, a single slider cannot produce a net flow because of time-reversal symmetry \citep{purcell1977life, ishimoto2012coordinate}.
Although each individual unit does not break time-reversal symmetry, a collection of sliders can still generate net flow by breaking this symmetry collectively \citep{khaderi2012fluid,wang2024programmable}.
In the standard convention, also used in the previous section, symplectic and antiplectic coordination are distinguished by comparing the direction of the effective stroke with that of phase propagation.
Since an isolated slider is reciprocal and has no intrinsic effective stroke, we define the coordination type based on the direction of phase propagation relative to the resulting fluid transport direction.

The model is studied using two complementary approaches: analytical and numerical methods.
We first describe the numerical method used to solve the slider dynamics, while the asymptotic solution is derived using a regular perturbation method in the following section.
In the numerical analysis, the slider dynamics are solved using the force \eqref{eq:slider_force} and the mobility relation \eqref{eq:bead_velocity}.
The slider velocity $\dot{s}_i = \bm{e}_s \cdot \bm{v}_i$ is obtained from the mobility relation together with the constraint $\bm{e}_{s,\perp} \cdot \bm{v}_i = 0$, where $\bm{v}_i$ is the bead velocity. 
The position $s_i$ is updated using a forward Euler scheme with time step \(\Delta t^*=4.0\times10^{-5}\).
We sweep the phase difference \(\delta\in[-\pi,\pi]\) and obtain \(\Delta=\operatorname*{arg\,max}_{\delta\in[-\pi,\pi]}Q^*(\delta)\), with the corresponding maximised flow rate \(Q_{\max}^*=Q^*(\Delta)\).

\subsection{Regular perturbation analysis in the weak-coupling limit}\label{sec:weak_coupling}

In this section, we derive an asymptotic description of the flow generation using the regular perturbation method, and analyse the optimal phase difference $\Delta$ for given $(\ell^*, \phi)$ conditions.
The following three approximations are introduced to make the system analytically tractable.
First, we consider the far-field limit $\ell^* \gg 1$, in which the hydrodynamic interaction between the two sliders is weak compared with the self-mobility.
Based on the far-field expansion of the Blake tensor \citep{blake1971note} shown in Appendix \ref{app:farfield_pair_mobility}, we retained only the leading-order terms of the hydrodynamic coupling, up to $O(\ell^{-4})$.
Second, the wall correction to the self-mobility is omitted, and the self-mobility is approximated as a constant, $\gamma_0^*=1/(6\pi a^*)$.
Third, the constraint force $\Lambda_i$ is neglected because it contributes only at $O((\ell^*)^{-6})$.

\subsubsection{Weak-coupling reduction}

To simplify the projected dynamics, we introduce the horizontal projections of the slider displacements, $\xi_i^* \coloneqq s_i^*\cos\phi$ \((i=1,2)\),  together with the sum and difference variables $\xi_+^* \coloneqq \xi_1^*+\xi_2^*$ and $\xi_-^* \coloneqq \xi_2^*-\xi_1^*$.
Here, \(\xi_+^*/2\) is the horizontal shift of the centroid,
while \(\xi_-^*\) is the instantaneous difference of the displacement.
The bead positions are then written as
\begin{equation}
x_1^*=-\frac{\ell^*}{2}+\xi_1^*,\qquad
z_1^*=1-\xi_1^*\tan\phi,
\qquad
x_2^*=\frac{\ell^*}{2}+\xi_2^*,\qquad
z_2^*=1-\xi_2^*\tan\phi.
\label{eq:slider_geometry}
\end{equation}
Using the far-field expansion of the pair mobility
\begin{equation}
\mathcal C^*(t^*)
\coloneqq
\frac{3}{2\pi}z_1^*z_2^*
\left[
\frac{\cos^2\phi}{(\ell^*+\xi_-^*)^3}
+
\frac{\sin^2\phi\,\xi_-^*}{(\ell^*+\xi_-^*)^4}
\right].
\label{eq:Cstar_app}
\end{equation}
derived in Appendix~\ref{app:farfield_pair_mobility}, the dynamics reduce to
\begin{equation}
\begin{split}
\dot{\xi}_+^*
&=
\gamma_0^*
\left(
1+\frac{\mathcal{C}^*}{\gamma_0^*}
\right)
\bigl(f_+^*\cos\phi-k_r^*\xi_+^*\bigr),
\\
\dot{\xi}_-^*
&=
\gamma_0^*
\left(
1-\frac{\mathcal{C}^*}{\gamma_0^*}
\right)
\bigl(f_-^*\cos\phi-k_r^*\xi_-^*\bigr),
\end{split}
\label{eq:xi_final}
\end{equation}
where \(f_1^*(t^*)=\cos(\omega^* t^*+\delta)\), \(f_2^*(t^*)=\cos(\omega^* t^*)\), \(f_+^*\coloneqq f_1^*+f_2^*\), and \(f_-^*\coloneqq f_2^*-f_1^*\).
Following the usual scaling used in minimal descriptions of hydrodynamically coupled ciliary rotors \citep{uchida2011generic,uchida2012hydrodynamic}, we identify the weak-coupling parameter with the dimensionless strength of the wall-mediated pair mobility relative to the isolated self-mobility.
Since \(z_1^*,z_2^*=O(1)\) and \(|\xi_-^*|\ll \ell^*\) in the weak-coupling regime, the projected pair-mobility coefficient satisfies \(\mathcal{C}^*=O(\ell^{*-3})\),
and hence,
\begin{equation}
\frac{1}{\gamma_0^*}\sup_{t^*\in[0,T^*]}
\left|\mathcal{C}^*(t^*)\right|
=
O\!\left(\frac{a^*}{\ell^{*3}}\right),
\end{equation}
where the leading contribution to \(\mathcal C^*/\gamma_0^*\) has the scale
\(9a^*/\ell^{*3}\).
We therefore define
\begin{equation}
\varepsilon \coloneqq \frac{9a^*}{\ell^{*3}}\ll1
\end{equation}
as a representative measure of the far-field hydrodynamic coupling strength.
We now expand the variable $\xi_\pm^*$ as
\begin{equation}
\xi_\pm^*
=
\xi_{\pm,0}^*
+
\xi_{\pm,1}^*
+
\xi_{\pm,2}^*
+\cdots,
\qquad
\xi_{\pm,n}^* = O(\varepsilon^n),
\label{eq:regular_expansion_star}
\end{equation}
where the powers of \(\varepsilon\) are absorbed into the correction terms themselves.
Here \(\xi_{\pm,0}^*\) denotes the isolated-slider response, and \(\xi_{\pm,1}^*\) is the actual \(O(\varepsilon)\) correction induced by the pair interaction.
This expansion provides the basis for identifying the leading phase-dependent correction to the flow rate.

\subsubsection{Flow-rate expression}
Using \eqref{eq:qx_reward}, the instantaneous flow rate generated by the two sliders is
\begin{equation}
q^*(t^*)
=
\frac{1}{\pi}
\sum_{i=1}^2
z_i^*
\bigl(f_i^*\cos\phi-k_r^*\xi_i^*\bigr),
\end{equation}
and the cycle-averaged dimensionless flow rate is therefore
\begin{equation}
Q^*
=
\frac{1}{T^*}\int_0^{T^*} q^*(t^*)\,\mathrm{d}t^*.
\end{equation}
Substituting \(z_i^*=1-\xi_i^*\tan\phi\) and rewriting the result in terms of \(\xi_\pm^*\) and \(f_\pm^*\), we obtain
\begin{equation}
Q^*
=
\frac{1}{\pi T^*}
\int_0^{T^*}
\left[
-k_r^* \xi_+^*
-\frac{\sin\phi}{2}
\left(
\xi_+^* f_+^* 
+\xi_-^* f_-^* 
\right)
+\frac{k_r^*\tan\phi}{2}
\left(
\xi_+^{*2}+\xi_-^{*2}
\right)
\right]\mathrm{d}t^*,
\label{eq:Qstar_pm1}
\end{equation}
where we have used \(\int_0^{T^*}f_+^*(t^*)\,\mathrm{d}t^*=0\).
Defining \(\Gamma_\pm^*(t^*) \coloneqq \gamma_0^* \pm \mathcal{C}^*(t^*)\), \eqref{eq:xi_final} becomes
\begin{equation}
f_\pm^* \cos\phi-k_r^*\xi_\pm^*
=
\frac{\dot{\xi}_\pm^*}{\Gamma_\pm^*}.
\label{eq:fpm_relation}
\end{equation}
Substituting \eqref{eq:fpm_relation} into \eqref{eq:Qstar_pm1} yields
\begin{equation}
\begin{split}
Q^*
&=
\frac{1}{\pi T^*}
\int_0^{T^*}
\left[
-k_r^* \xi_+^*
-\frac{\tan\phi}{2}
\left(
\frac{\xi_+^* \dot{\xi}_+^*}{\Gamma_+^*}
+
\frac{\xi_-^* \dot{\xi}_-^*}{\Gamma_-^*}
\right)
\right]\mathrm{d}t^* \\
&=
-
\frac{k_r^*}{\pi}
\langle \xi_+^* \rangle
-\frac{\tan\phi}{2 \pi}
\left \langle
\frac{\xi_+^* \dot{\xi}_+^*}{\Gamma_+^*}
+
\frac{\xi_-^* \dot{\xi}_-^*}{\Gamma_-^*}
\right \rangle
\end{split}
\end{equation}
where $\langle \cdot \rangle$ denotes the cycle average over one period $T^*$.
Since
\begin{equation}
\left\langle \frac{\xi_\pm^*\dot{\xi}_\pm^*}{\Gamma_\pm^*} \right\rangle
=
\frac{1}{2}
\left\langle
\frac{d}{\mathrm{d}t^*}
\left(
\frac{\xi_\pm^{*2}}{\Gamma_\pm^*}
\right)
\right\rangle
+
\frac{1}{2}\,
\left\langle
\left(\frac{\xi_\pm^*}{\Gamma_\pm^*} \right)^2 \,
\dot{\Gamma}_\pm^*
\right\rangle
=
\frac{1}{2}\,
\left\langle
\left(\frac{\xi_\pm^*}{\Gamma_\pm^*} \right)^2 \,
\dot{\Gamma}_\pm^*
\right\rangle
,
\end{equation}
where the last equality follows from the periodicity of \(\xi_\pm^*\) and \(\Gamma_\pm^*\), we finally obtain, using \(\dot{\Gamma}_\pm^*(t^*)=\pm \dot{\mathcal C}^*(t^*)\),
\begin{equation}
Q^*
=
-
\frac{k_r^*}{\pi}
\langle \xi_+^* \rangle
-\frac{\tan\phi}{4 \pi}
\left \langle
\dot{\mathcal C}^*
\left(
\frac{\xi_+^{*2}}{\Gamma_+^{*2}}
-
\frac{\xi_-^{*2}}{\Gamma_-^{*2}}
\right)
\right \rangle .
\label{eq:Qstar_compact}
\end{equation}
Equation~\eqref{eq:Qstar_compact} gives an exact decomposition of the cycle-averaged flow rate.
The first term represents the flow generated by the shift in the time-averaged position \(\langle \xi_+^* \rangle/2\), as described in detail later.
The second term is a correction arising from the time-dependent pair mobility.
Note that the second term is subdominant because its dominant Fourier component, located at the angular frequency $\omega$, contributes only partially to the zero-frequency projection.
This can be understood from the orthogonality of the Fourier components: for the harmonic leading-order motion, $\xi_{+}^{*2}-\xi_{-}^{*2}$ contains only a mean component and a second harmonic, whereas the dominant component of $\dot{\mathcal C}^*$ is a first harmonic.
In this estimate, we use $(\Gamma_\pm^*)^{-2}\simeq(\gamma_0^*)^{-2}$, assuming that the pair contribution $\mathcal C^*$ is small relative to $\gamma_0^*$.

\subsubsection{Zeroth-order response}
Since \(\mathcal{C}^* \sim O(\varepsilon)\), substituting the regular expansion \eqref{eq:regular_expansion_star} into \eqref{eq:xi_final} and collecting the \(O(1)\) terms yields
\begin{equation}
\dot{\xi}_{\pm,0}^*+\gamma_0^*k_r^* \xi_{\pm,0}^*=
\gamma_0^* f_\pm^*\cos\phi.
\label{eq:xi0_pm}
\end{equation}
Thus, at leading-order, the two sliders are decoupled and each mode behaves as a forced first-order relaxation process.
From \eqref{eq:xi0_pm}, the forcing terms are
\begin{equation}
f_+^*(t^*)=2\cos\!\left(\frac{\delta}{2}\right)\cos\!\left(\omega^* t^*+\frac{\delta}{2}\right),
\qquad
f_-^*(t^*)=2\sin\!\left(\frac{\delta}{2}\right)\sin\!\left(\omega^* t^*+\frac{\delta}{2}\right).
\label{eq:fpm_explicit}
\end{equation}
The corresponding steady periodic solutions are
\begin{equation}
\xi_{+,0}^*(t^*)=A^*\cos\!\left(\frac{\delta}{2}\right)\cos\chi^*,
\qquad
\xi_{-,0}^*(t^*)=A^*\sin\!\left(\frac{\delta}{2}\right)\sin\chi^*
\label{eq:xi0_sol}
\end{equation}
where
\begin{equation}
\chi^*
\coloneqq
\omega^* t^*+\frac{\delta}{2}
-\arctan\!\left(\frac{\omega^*}{\gamma_0^*k_r^*}\right),
\qquad
A^*
\coloneqq
\frac{2\gamma_0^*\cos\phi}
{\sqrt{(\gamma_0^*k_r^*)^2+\omega^{*2}}}
\label{eq:Astar_chistar}
\end{equation}
and \(A^*\) is the amplitude of the zeroth-order horizontal slider motion.

The \(O(1)\) contribution to the cycle-averaged flow rate follows from \eqref{eq:Qstar_compact}. Since the interaction-dependent term in \eqref{eq:Qstar_compact} is itself \(O(\varepsilon)\), one finds
\begin{equation}
Q_0^*
=
-\frac{k_r^*}{\pi} \langle \xi_{+,0}^* \rangle.
\label{eq:Q0_def}
\end{equation}
Because \(\xi_{+,0}^*\) is a zero-mean periodic function, we obtain \(Q_0^*=0\).
This result reflects the fact that, at \(O(1)\), the hydrodynamic interaction is absent and each slider executes an independent reciprocal oscillation.
Accordingly, the leading-order motion is kinematically reversible and generates no cycle-averaged transport, as expected from the reversibility principle underlying the scallop theorem at low Reynolds numbers \citep{purcell1977life,ishimoto2012coordinate}.

\begin{figure}
    \centering
    \includegraphics[width=0.80\linewidth]{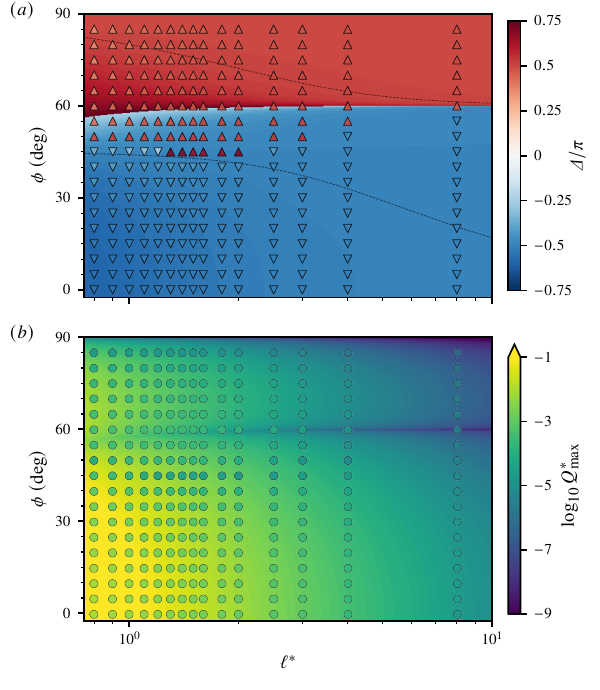}
\caption{
Phase-dependent flow generation in the reduced tilted-slider model under conditions \(a^*=0.05\), \(k_r^*=2\), and \(\omega^*=\pi\).
(a) Flow-maximising phase difference \(\Delta/\pi\) for different \((\ell^*,\phi)\).
The background contour map shows the values obtained from the perturbation analysis \eqref{eq:Q1_final}, while the symbols show those obtained from phase sweeps using numerical simulations.
The symbol shapes, $\triangle$ and $\triangledown$, represent symplectic and antiplectic coordination, respectively.
Dashed lines indicate the contours \(\Delta/\pi=\pm0.5\).
(b) Maximum cycle-averaged flow rate \(Q_{\max}^*\) for different \((\ell^*,\phi)\).
Colours represents the value \(\log_{10}Q_{\max}^*\).
\label{fig:fig6}
}
\end{figure}

\subsubsection{First-order phase-dependent flow rate}
We now turn to the first non-trivial contribution to the cycle-averaged transport.
Since the \(O(\varepsilon)\) correction to \eqref{eq:Qstar_compact} depends on the mean shift of
\(\xi_+^*\), it is sufficient here to consider the equation for \(\xi_{+,1}^*\).
Retaining terms $O(\varepsilon)$ in \eqref{eq:xi_final} gives
\begin{equation}
\dot{\xi}_{+,1}^*+\gamma_0^*k_r^*\,\xi_{+,1}^*
=
\mathcal{C}^*
\bigl(f_+^*\cos\phi-k_r^*\xi_{+,0}^*\bigr)
=
\frac{\mathcal{C}^*}{\gamma_0^*}\,\dot{\xi}_{+,0}^*,
\label{eq:xip1_rewrite}
\end{equation}
where the second equality follows from \eqref{eq:xi0_pm}.
To extract the contribution of \(\xi_{+,1}^*\) to the cycle-averaged flow rate, we average the \(+\) equation over one period.
Since \(\xi_{+,1}^*\) is periodic in the steady state, \(\langle \dot{\xi}_{+,1}^* \rangle = 0\), and we obtain
\begin{equation}
\langle \xi_{+,1}^* \rangle
= \frac{1}{k_r^*(\gamma_0^*)^2}
\left\langle \mathcal{C}^*\,\dot{\xi}_{+,0}^* \right\rangle
= -\frac{1}{k_r^*(\gamma_0^*)^2}
\left\langle \dot{\mathcal{C}}^*\,\xi_{+,0}^* \right\rangle.
\label{eq:mean_xi1_ibp}
\end{equation}
The \(O(\varepsilon)\) contribution to the flow rate follows from \eqref{eq:Qstar_compact}, and it can be written using \(\Gamma_\pm^{*-2}(t^*)=(\gamma_0^*)^{-2}(1+O(\varepsilon))\) as
\begin{equation}
Q_{\varepsilon}^*
=
-\frac{k_r^*}{\pi}\left\langle \xi_{+,1}^* \right\rangle
-\frac{\tan\phi}{4\pi(\gamma_0^*)^2}
\left\langle
\dot{\mathcal{C}}^*
\left(\xi_{+,0}^{*2}-\xi_{-,0}^{*2}\right)
\right\rangle,
\label{eq:Q1_pre_sub}
\end{equation}
where the first term arises from the mean \(O(\varepsilon)\) shift of \(\xi_+^*\), and the second from the \(O(\varepsilon)\) geometric condition.
Substituting \eqref{eq:mean_xi1_ibp}, and defining
\begin{equation}
W_0^*(t^*)
\coloneqq
\xi_{+,0}^*
-\frac{\tan\phi}{4}\left(\xi_{+,0}^{*2}-\xi_{-,0}^{*2}\right),
\label{eq:W0_def}
\end{equation}
we obtain
\begin{equation}
Q_{\varepsilon}^*
=
\frac{1}{\pi(\gamma_0^*)^2}
\left\langle
\dot{\mathcal{C}}^*\,W_0^*
\right\rangle
=
-\frac{1}{\pi(\gamma_0^*)^2}
\left\langle
\mathcal{C}^*\,\dot{W}_0^*
\right\rangle.
\label{eq:Q1_by_parts}
\end{equation}
Substituting the leading-order solution \eqref{eq:xi0_sol} into \(W_0^*\) and \(\mathcal{C}^*\), the period average in \eqref{eq:Q1_by_parts} can still be evaluated analytically.
Since the intermediate algebra is straightforward but lengthy, the explicit reduction is deferred to Appendix~\ref{app:first_order_flow}.
This yields
\begin{equation}
Q_{\varepsilon}^* (\delta)
=
-\frac{3\omega^*\cos^2\phi}{4\pi^3(\gamma_0^*)^2}
\left[
H_1^* \sin\delta
+
H_2^* \sin 2\delta
+
O\!\left(\frac{A^{*6}}{\ell^{*8}}\right)
\right].
\label{eq:Q1_final}
\end{equation}
Here
\begin{equation}
H_1^*(\ell^*, \phi)
=
\frac{\pi A^{*2}}{2\ell^{*4}}\bigl(3-\tan^2\phi\bigr)
+
\frac{\pi A^{*4}}{32\ell^{*6}}
\left[
\ell^{*2}\tan^2\phi\,\bigl(3-\tan^2\phi\bigr)
+60\bigl(1-\tan^2\phi\bigr)
\right]
\label{eq:H1_def}
\end{equation}
and
\begin{equation}
H_2^*(\ell^*, \phi)
=
\frac{\pi A^{*4}}{32\ell^{*6}}
\left[
\ell^{*2}\tan^2\phi\,\bigl(3-\tan^2\phi\bigr)
-30\bigl(1-\tan^2\phi\bigr)
\right].
\label{eq:H2_def}
\end{equation}
The candidates for the optimal phase difference $\Delta$ can be identified by solving the stationarity condition
\begin{equation}
\frac{\partial Q_{\varepsilon}^*}{\partial \delta}
=
H_1^* \cos\delta
+
2 H_2^* \cos 2\delta
=0,
\label{eq:dQ1_ddelta}
\end{equation}
and we finally obtain $\Delta (\ell^*, \phi)$. 

The leading-order flow generation is $Q^*_\varepsilon \propto \omega^* A^{*2} / \ell^{*4}$ and this quadratic scaling with the beating amplitude is in agreement with the previous studies \citep{taylor1951analysis,blake1971spherical,khaderi2012fluid}.
For a fixed forcing amplitude, the leading-order flow rate is maximised when the driving frequency matches the elastic relaxation rate, \(\omega^*=\gamma_0^*k_r^*\).
The existence of an optimal beating frequency can again be understood in the context of the $\mathrm{Sp}$ number, and the beating period has to be comparable with the relaxation time of the slider.
It is also important to note that the generated flow $Q^*_\varepsilon$ scales as $O(\ell^{-4})$, whereas the flow enhancement in Fig.~\ref{fig:fig4}(c) decays as $O(\ell^{-3})$.
This difference arises from the difference in the beating orbit: the tilted-slider model gives the $O(\ell^{-4})$ scaling because the effective--recovery stroke asymmetry is not retained.
Nevertheless, the tilted-slider model is sufficient to capture the main trends in the flow-maximising coordination, as shown in the next section.

\subsection{Flow-maximising coordination in the tilted-slider model}

We now analyse the flow generation of the tilted-slider model.
Figure~\ref{fig:fig6}(a) compares the flow-maximising phase difference $\Delta(\ell^*, \phi)$ obtained by the perturbation analysis and the numerical simulation, under the conditions $a^* = 0.05$, $k_r^* = 2$, and $\omega^* = \pi$; the background contour map represents the perturbation analysis, while the symbols show the numerical simulations described in \S\ref{subsec:tilted-slider-model}.
The two approaches lead to the same finding, interestingly showing that the flow-maximising coordination depends on the slider angle $\phi$.
They suggest that flow-maximising coordination is not universally antiplectic, and the sliders prefer antiplectic coordination when the beating angle is closer to the parallel configuration $\phi = 0$, while they prefer symplectic coordination when the angle is closer to the perpendicular configuration $\phi = \pi/2$.
Although the symplectic--antiplectic transition differs slightly between the two approaches in the near field, $\ell^* < 4$, their qualitative agreement suggests that the assumptions underlying the perturbation analysis were reasonable to capture the main trends.
Figure~\ref{fig:fig6}(b) shows the corresponding maximum flow rate $Q_{\max}^*$.
As expected, the largest flow generation is obtained at $\phi = 0$ and the smallest inter-slider spacing $\ell^*$.
For a fixed $\phi$, the flow decreases monotonically with increasing inter-slider spacing because the hydrodynamic coupling decays with distance.
The dependence on $\phi$, however, is not simply monotonic: the flow exhibits two local maxima in the symplectic and antiplectic regimes, and becomes small near the transition between the two states.

\begin{figure}
    \includegraphics[width=0.7\linewidth]{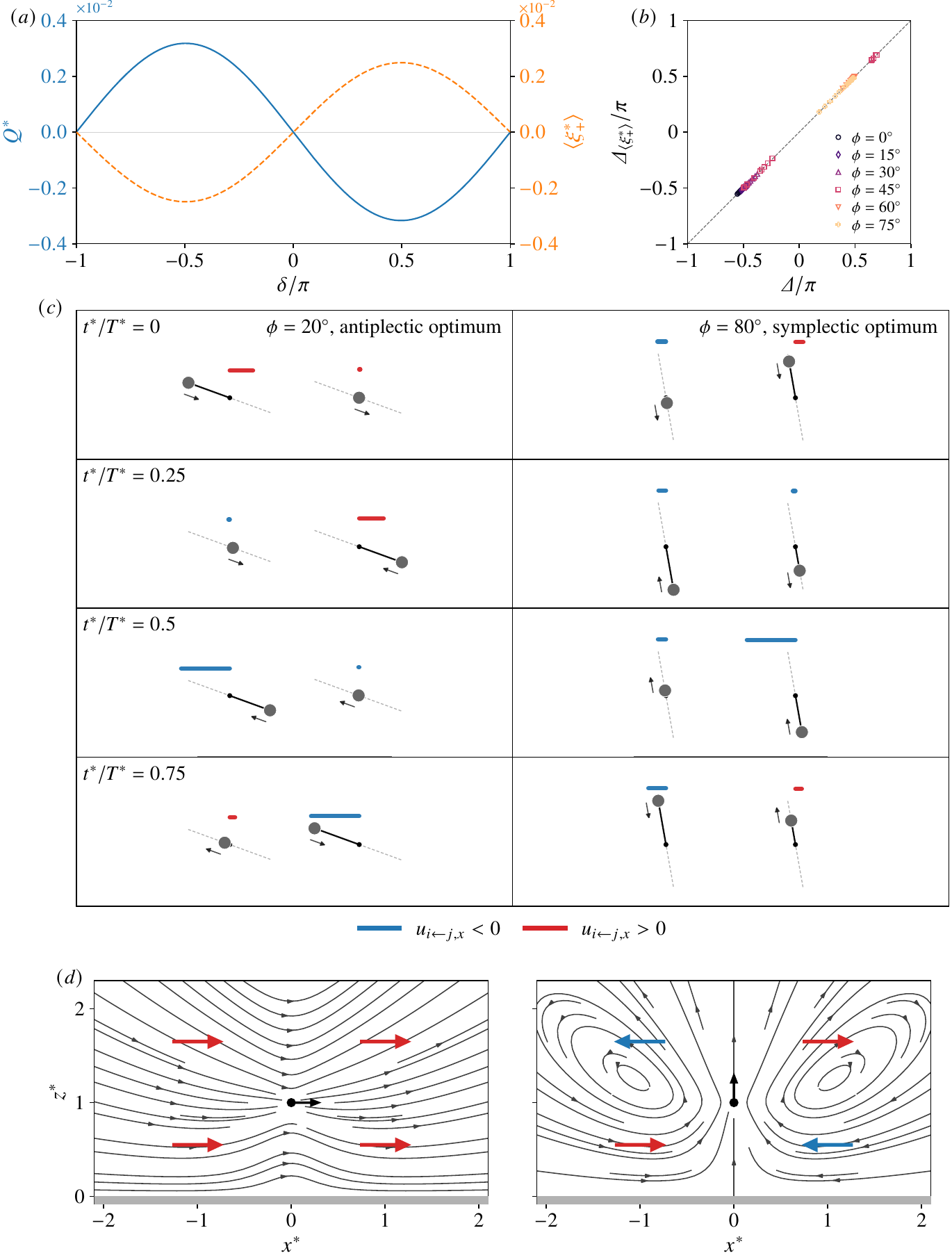}
\caption{
Mechanism of flow generation in the tilted-slider model.
(a) Phase-difference $\delta$ dependence of the flow rate $Q^*$ and the time-averaged slider position $\langle  \xi_+^* \rangle$, obtained from numerical simulations under the conditions $\ell^*=1.0$ and $\phi=20^\circ$.
The flow rate increases as the position becomes more negative.
(b) Comparison between the flow-maximising phase difference \(\Delta\) and the phase difference \(\Delta_{\langle \xi_+^*\rangle}\) that minimises \(\langle \xi_+^*\rangle\), obtained from numerical simulations over different \(\ell^*\) and \(\phi\).
The dashed line indicates \(\Delta_{\langle \xi_+^*\rangle}=\Delta\).
(c) Snapshots over one cycle from numerical simulations for the antiplectic optimum at $\phi=20^\circ$ and the symplectic optimum at $\phi=80^\circ$.
Blue and red bars show the $x$-component of the velocity induced by pair mobility, $u_{i \leftarrow j, x}$; the bar length represents its magnitude, and the colour denotes the sign of $u_{i \leftarrow j, x}$, with red and blue indicating positive and negative values, respectively.
Note that the bar lengths are normalised by the maximum magnitude of $u_{i \leftarrow j, x}$ evaluated at the four snapshot times.
(d) Wall-mediated flow generated by horizontal and vertical point forces near a no-slip wall.
\label{fig:fig7}
}
\end{figure}

The origin of the flow generation can be understood from \eqref{eq:Qstar_compact}.
Since the second term of \eqref{eq:Qstar_compact} is small relative to the first, as discussed in the previous section, the flow generation is primarily determined by the first term, $-k_r^* \langle \xi_+^* \rangle/\pi$.
Thus, maximising the flow is equivalent to driving the time-averaged horizontal position of the sliders, $\langle \xi_+^* \rangle$, further in the negative direction.
Note that this negative shift can arise only through hydrodynamic coupling between the sliders, since the self-mobility alone would produce only symmetric oscillations about the origin.
The negative shift enables the sliders to generate flow by introducing an asymmetry over the beating cycle.
When the time-averaged position is shifted in the $-x$ direction, the elastic spring assists the $+x$ stroke and enhances its effective forcing, whereas it resists the $-x$ stroke and reduces the forcing.
Figure~\ref{fig:fig7}(a) provides a visual confirmation of this relationship using the numerical simulations, showing that the obtained flow $Q^*$ is negatively correlated with $\langle \xi_+^* \rangle$ and that larger flow is generated for smaller values of $\langle \xi_+^* \rangle$.
Figure~\ref{fig:fig7}(b) also confirms that the phase difference that maximises the flow, $\Delta$, agrees well with the phase difference $\Delta_{\langle \xi_+^* \rangle}$ that minimises $\langle \xi_+^* \rangle$ over a wide range of $\ell^*$ and $\phi$.

Having established that a negative shift of the time-averaged ciliary position contributes to the flow generation, we next ask why the optimal coordination depends on the beating angle $\phi$.
Figure~\ref{fig:fig7}(c) shows time-series snapshots of the optimal coordination for two different beating angles, $\phi = 20^\circ$ and $80^\circ$.
The blue and red bars in the snapshots indicate the negative and positive $x$-velocity induced by the pair mobility, $u_{i \leftarrow j, x}$, respectively.
We first take a closer look at the left column, which corresponds to the case of a small angle $\phi$.
In this regime, antiplectic coordination maximises the negative shift of the time-averaged position through the asymmetry of the hydrodynamic coupling.
During the recovery stroke ($0.5 < t^*/T^* < 0.75$; third to fourth row), both sliders move in the $-x$ direction, and the pair mobility $G_{11}$ contributes to the negative shift in $\langle \xi_+^*\rangle$, where $\bm{G}$ is the Blake tensor \citep{blake1971note}.
During the effective stroke ($0 < t^*/T^* < 0.25$; first to second row), by contrast, both sliders move in the $+x$ direction, and the pair mobility $G_{11}$ contributes to the positive shift.
Since the hydrodynamic coupling is stronger at shorter separations, the negative contribution during the recovery stroke exceeds the positive contribution during the effective stroke, resulting in a net negative shift in the time-averaged position.
This shorter separation during the recovery stroke is consistent with the shielding/obstruction picture proposed in previous studies.
Although both our analysis and previous studies identify antiplectic coordination as optimal, the mechanistic interpretation is somewhat different.
The shielding/obstruction picture views the enhancement in terms of competing flow fields generated by neighbouring cilia, whereas our analysis views it in terms of how asymmetric pair mobility shifts the time-averaged position and thereby modifies the elastic response.
These explanations are not mutually exclusive, and both mechanisms may contribute to the same antiplectic advantage.
Next, we focus on the right column, corresponding to the case with large $\phi$, where symplectic coordination is optimal.
Because the stroke is oriented primarily in the $z$-direction in this case, the hydrodynamic coupling is mainly governed by the $G_{13}$ component, with an additional contribution from $G_{33}$ in the near field.
When the sliders move in the $+z$ direction, corresponding to the recovery stroke, the resulting flow field is shown in Fig.~\ref{fig:fig7}(d).
Since this two-vortex flow field exhibits a negative $x$-velocity in the upper-left and lower-right regions of the figure, the two sliders should adopt a ``\textbackslash"-shaped diagonal arrangement during the recovery stroke to maximise the negative shift.
This arrangement, with the rear slider positioned above the front slider, can be seen in the third and fourth rows of the figure.
When the sliders move in the $-z$ direction, corresponding to the effective stroke, the flow-field pattern is reversed, with negative $x$-velocity appearing in the upper-right and lower-left regions of the figure.
As a result, a ``/"-shaped diagonal arrangement maximises the negative shift during the effective stroke, as seen in the first and second rows of the figure.
Since these diagonal arrangements can be achieved only with a symplectic phase difference, symplectic coordination is preferred at these angles.

Note that this slider model is related to previous studies on symmetrically beating artificial cilia \citep{khaderi2012fluid,wang2024programmable}, in which flow is generated by collectively breaking time-reversal symmetry despite each individual unit undergoing reciprocal motion.
These studies showed, numerically \citep{khaderi2012fluid} and experimentally \citep{wang2024programmable}, that the resulting flow is directed opposite to the direction of phase propagation, consistent with the present result with $\phi = 0$.
Although this point was not examined in those studies, it would be interesting to test whether a similar negative shift in the time-averaged position also occurs in these symmetrically beating systems.

\begin{figure}
    \centering
    \includegraphics[width=0.8\linewidth]{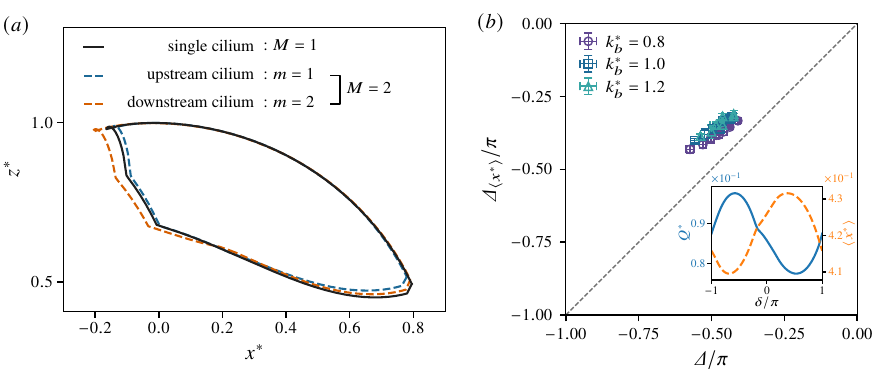}
\caption{
Negative shift in the time-averaged position for two cilia system.
(a) Comparison of the tip-bead trajectories for the single cilium ($M=1$) and two cilia ($M=2$) systems, for  $\ell^*=1.3$ and $k_b^*=0.8$.
Compared with the single cilium case, the two cilia exhibit a negative shift in the time-averaged position relative to the effective-stroke direction.
(b) Comparison between the flow-maximising phase difference \(\Delta\) and the phase difference \(\Delta_{\langle x^*\rangle}\) that minimises \(\langle x^*\rangle\).
Different symbols denote different bending stiffnesses \(k_b^*\) while varying the inter-cilium distance \(\ell^*\).
The dashed line indicates \(\Delta_{\langle x^*\rangle}=\Delta\).
Inset: Representative example showing the relation between the flow rate $Q^*$ and the time-averaged tip-bead position $\langle x^* \rangle$ when varying the phase difference $\delta$. The conditions are $\ell^*=1.2$ and $k_b^*=1.0$.
\label{fig:fig8}
}
\end{figure}

\subsection{Negative shift in the cilia model}
The tilted-slider model shows that the optimal coordination corresponds to the phase difference that induces a negative shift in the time-averaged position, which explains why the optimal coordination depends on the beat geometry $\phi$.
We finally examine whether the negative shift in the time-averaged position is also observed in the two cilia system analysed in \S~\ref{subsec:twocilia}.
Figure~\ref{fig:fig8}(a) compares the tip-bead trajectories for the single cilium ($M=1$) and two cilia ($M=2$) systems.
Note that the trajectory for the two cilia system is obtained from the flow-maximising coordination identified by the phase-sweep analysis.
The figure shows that a negative shift in the time-averaged position indeed appears in the two cilia system.
The downstream cilium ($m=2$) exhibits a larger negative shift, consistent with the observation that the downstream cilium contributes more strongly to the flow enhancement (data not shown).
The inset of figure~\ref{fig:fig8}(b) shows the flow rate $Q^*$ and the corresponding horizontal shift $\langle x^*\rangle$ as functions of the imposed phase difference $\delta$, under the conditions $\ell^*=1.2$ and $k_b^*=1.0$.
As seen in the tilted-slider model, the flow rate is negatively correlated with the horizontal shift, and larger flow rates are expected when the time-averaged position is shifted in the negative direction.
The two cilia prefer antiplectic coordination, $\Delta < 0$, and this preference is consistent with the prediction of the tilted-slider model for small beating angles $\phi$.
Figure~\ref{fig:fig8}(b) compares the flow-maximising phase difference $\Delta$ with the phase difference $\Delta_{\langle x^* \rangle}$ that minimises $\langle x^* \rangle$ over all simulated conditions, showing again that the negative shift provides a useful criterion for predicting the flow-maximising coordination.
In Appendix~\ref{app:shear_phase_selection}, we briefly examine how the optimal coordination is altered by externally modifying the beating orbit with a background simple shear flow, and show that the resulting phase-selection trend is consistent with the prediction of the tilted-slider model.

The above analysis shows that collective flow enhancement is associated with the negative shift induced by hydrodynamic coupling with neighbouring cilia.
This shift enhances transport through its coupling with ciliary elasticity: the elastic restoring force assists the effective stroke, whereas it resists the recovery stroke.
These results, therefore, highlight the importance of an elastohydrodynamic perspective in understanding flow enhancement by metachronal waves.

\section{Conclusions}\label{sec:conclusions}

In this study, we analysed flow-maximising coordination in cilia arrays, focusing on the mechanism that selects the optimal coordination and its dependence on beat geometry.
We first used reinforcement learning to identify optimal coordination in a bead--spring cilia model.
Across the explored range of inter-cilium spacing, bending stiffness, and cilia number, reinforcement learning selected antiplectic coordination as the flow-maximising state.
Phase-sweep analysis further showed that the flow enhancement can be explained mainly by the phase difference between neighbouring cilia.
Motivated by these findings, we built the tilted-slider model that retains the driving force, the elastic restoring force, and the hydrodynamic coupling between the sliders.
Regular perturbation analysis showed that flow-maximising coordination is characterised by a shift of the time-averaged slider position opposite to the effective stroke direction.
This negative shift enhances flow generation by introducing an asymmetry over the beating cycle: the elastic restoring force assists the effective stroke, whereas it resists the recovery stroke.
The flow-maximising coordination depends on beat geometry, and some beat geometries even favour symplectic coordination.
Overall, these results show that flow enhancement is not governed by hydrodynamic coupling alone, but emerges from its interplay with ciliary elasticity, highlighting the importance of elastohydrodynamics in flow-maximising coordination.
The tilted-slider model helps rationalise how flow-maximising coordination changes with inter-cilium spacing, ciliary elasticity, and beat geometry, and this parameter dependence may be useful for designing efficient artificial ciliary systems.
Although autonomous biological cilia and externally actuated cilia differ in how their motions are determined, the negative shift identified here may also provide a useful perspective for understanding efficient fluid transport in biological ciliary arrays.
We expect that the insights obtained here may provide useful guidance for future efforts to understand and design efficient ciliary transport systems.


\appendix

\section{Estimate for the weighting factor $\alpha$} \label{app:alpha}

To choose the weighting factor $\alpha$, we compare the overdamped relaxation times of the basal and middle joints in a simplified local estimate for a single cilium.
Hydrodynamic interactions and extensional compliance are neglected, both segments are treated as rigid links of length $l_c=L/2$, and the rotational drag coefficient is taken as $\zeta_\theta = 6\pi\mu a l_c^2$.
Let $\psi_{01}$ and $\psi_{12}$ denote the absolute segment angles, so that $\theta_0=\psi_{01}$ and $\theta_1=\psi_{12}-\psi_{01}$.
We compare the two joint responses under the same constant maximal torque, $\tau_0=\tau_1=\tau_{\max}$.
For the basal joint, the overdamped angular balance is
\begin{equation}
\zeta_\theta \dot{\theta}_0=\alpha\left(\tau_0-k_b\theta_0\right),
\end{equation}
so that the relaxation time is
\begin{equation}
T_{\mathrm{relax},0}=\frac{\zeta_\theta}{\alpha k_b}.
\end{equation}
Because $\alpha$ multiplies both the elastic restoring torque and the active torque, the steady relation remains $\theta_0^\infty=\tau_0/k_b$.
For the middle joint, the corresponding local balances are $\zeta_\theta \dot{\psi}_{01}=-(\tau_1-k_b\theta_1)$ and $\zeta_\theta \dot{\psi}_{12}=\tau_1-k_b\theta_1$.
Subtracting these equations yields
\begin{equation}
\zeta_\theta \dot{\theta}_1=\zeta_\theta(\dot{\psi}_{12}-\dot{\psi}_{01})
=
2(\tau_1-k_b\theta_1),
\end{equation}
and hence
\begin{equation}
T_{\mathrm{relax},1}=\frac{\zeta_\theta}{2k_b}.
\end{equation}
Equating the two relaxation times, $T_{\mathrm{relax},0}=T_{\mathrm{relax},1}$, gives $\alpha=2$,
which is used in both the bending energy and the active work function throughout the simulations. 

\section{Far-field expression of the pair mobility for tilted-slider model}
\label{app:farfield_pair_mobility}

We derive the interaction coefficient $\mathcal C^*(t^*)$ appearing in \eqref{eq:xi_final} in this appendix.
We first obtain the far-field form of the relevant components of the Blake-tensor, and then specialise the projected pair mobility to the tilted-slider geometry.

\subsection{Far-field expansion of the Blake-tensor}
From the observation point \(\bm r=(x,0,z)\) and source point \(\bm r_0=(x_0,0,z_0)\), we define
\begin{equation}
\hat{x} \coloneqq x-x_0,
\qquad
r_\parallel \coloneqq |\hat{x}|.
\label{eq:parallel_distance_app}
\end{equation}
In the far-field regime, $\max\{z^*,z_0^*\}\ll r_\parallel^*$, the free-space Stokeslet and its image in the Blake representation cancel at $O((r_\parallel^*)^{-1})$, so that the leading wall-mediated contribution arises only at higher orders.
Expanding in powers of $z^*/r_\parallel^*$ and $z_0^*/r_\parallel^*$, one obtains
\begin{align}
G_{11}^*(\bm r^*,\bm r_0^*)
&=
\frac{3}{2\pi}\,
\frac{z^* z_0^*\,(\hat{x}^*)^{2}}{(r_\parallel^*)^{5}}
+O((r_\parallel^*)^{-5}),
\label{eq:G11_general_app}
\\
G_{13}^*(\bm r^*,\bm r_0^*)
&=
-\frac{3}{2\pi}\,
\frac{z^* (z_0^*)^{2}\,\hat{x}^*}{(r_\parallel^*)^{5}}
+O((r_\parallel^*)^{-6}),
\label{eq:G13_general_app}
\\
G_{31}^*(\bm r^*,\bm r_0^*)
&=
\frac{3}{2\pi}\,
\frac{(z^*)^{2}z_0^*\,\hat{x}^*}{(r_\parallel^*)^{5}}
+O((r_\parallel^*)^{-6}),
\label{eq:G31_general_app}
\end{align}
while
\begin{equation}
G_{33}^*(\bm r^*,\bm r_0^*)=O((r_\parallel^*)^{-5}).
\label{eq:G33_general_app}
\end{equation}
Note that the $y$-components of $\bm{G}$ were omitted here because they vanish and therefore do not contribute to the dynamics.
The scaling of these terms will be important below in obtaining the asymptotic expression of the pair mobility; $G_{11}^*$ has the lowest leading-order $O((r_\parallel^*)^{-3})$, followed by $G_{13}^*$ and $G_{31}^*$ with $O((r_\parallel^*)^{-4})$, and $G_{33}^*$ appears only at $O((r_\parallel^*)^{-5})$.

\subsection{Pair mobility in the tilted-slider geometry}

The tangential force acting on the $j$-th bead is
\begin{equation}
\bm F_j^{*,\parallel} =
\bigl(f_j^*-k_r^* s_j^* \bigr)\bm e_s
\end{equation}
where $\bm e_s=(\cos\phi,0,-\sin\phi)$ is the slider axis.
Note that the normal constraint force $\Lambda$, which appears in equation~\eqref{eq:slider_force}, is neglected since $\Lambda = O((\ell^*)^{-3})$ would contribute only an $O((\ell^*)^{-6})$ correction to the pair mobility, once we assume that the drift due to self mobility is negligible. 
The translational velocity of the $i$-th bead due to the hydrodynamic coupling with the $j$-th bead ($i\neq j$) is
\begin{equation}
\dot{\bm r}_{i \leftarrow j}^*
=
\bm{G}^*(\bm r_i^*,\bm r_j^*) \bm F_j^{*,\parallel}.
\end{equation}
By projecting the translational velocity onto the slider direction \citep[e.g.][]{uchida2011generic}, we obtain
\begin{equation}
\dot s_{i \leftarrow j}^*
=
\bm e_s\cdot \dot{\bm r}_{i \leftarrow j}^*
=
\widetilde{\mathcal C}^*(\bm r_i^*,\bm r_j^*)
\bigl(f_j^*-k_r^* s_j^*\bigr),
\label{eq:s_pair_contribution_app}
\end{equation}
where $\widetilde{\mathcal C}$ is the projected pair mobility defined as
\begin{equation}
\begin{split}
\widetilde{\mathcal C}^*(\bm r_i^*,\bm r_j^*)
&\coloneqq
\bm e_s \cdot \bm{G}^*(\bm r_i^*,\bm r_j^*)\cdot \bm e_s \\
&=
\cos^2\phi\,G_{11}^*(\bm r_i^*,\bm r_j^*)
-\sin\phi\cos\phi
\bigl[
G_{13}^*(\bm r_i^*,\bm r_j^*)
+
G_{31}^*(\bm r_i^*,\bm r_j^*)
\bigr]
+\sin^2\phi\,G_{33}^*(\bm r_i^*,\bm r_j^*).
\end{split}
\label{eq:Ctilde_def_app}
\end{equation}
From the definition $\xi_i^*=s_i^*\cos\phi$, we have
\begin{equation}
\dot\xi_{i \leftarrow j}^*
=
\dot{s}_{i \leftarrow j}^* \cdot \cos\phi
=
\widetilde{\mathcal C}^*(\bm r_i^*,\bm r_j^*)
\bigl(f_j^*\cos\phi-k_r^* \xi_j^*\bigr).
\label{eq:xi_pair_contribution_app}
\end{equation}
We now specialise in the present two-slider system and evaluate the projected pair mobility for the ordered pair $(i,j)=(1,2)$.
For the opposite ordering, Lorentz reciprocity gives $\bm{G}^*(\bm r_2^*,\bm r_1^*)=\{\bm{G}^*(\bm r_1^*,\bm r_2^*)\}^{\mathsf T}$, and hence $\widetilde{\mathcal C}^*(\bm r_2^*,\bm r_1^*)=\widetilde{\mathcal C}^*(\bm r_1^*,\bm r_2^*)$.
Using the slider geometry \eqref{eq:slider_geometry}, one has
\begin{equation}
\hat{x}^*=x_1^*-x_2^*=-(\ell^*+\xi_-^*),
\qquad
r_\parallel^*=\ell^*+\xi_-^*>0,
\qquad
z_2^*-z_1^*=-\xi_-^*\tan\phi.
\label{eq:pair_geometry_app}
\end{equation}
Substituting \eqref{eq:pair_geometry_app} into \eqref{eq:G11_general_app} gives
\begin{equation}
G_{11}^*(\bm r_1^*,\bm r_2^*)
=
\frac{3}{2\pi}\,
\frac{z_1^* z_2^*}{(\ell^*+\xi_-^*)^3}
+O((r_\parallel^*)^{-5}),
\label{eq:G11_pair_app}
\end{equation}
while combining \eqref{eq:G13_general_app} and \eqref{eq:G31_general_app} yields
\begin{equation}
G_{13}^*(\bm r_1^*,\bm r_2^*)+G_{31}^*(\bm r_1^*,\bm r_2^*)
=
\frac{3}{2\pi}\,
\frac{z_1^* z_2^*(z_2^*-z_1^*)}{(\ell^*+\xi_-^*)^4}
+O((r_\parallel^*)^{-6}).
\label{eq:G13G31_pair_app}
\end{equation}
Substituting \eqref{eq:G11_pair_app} and \eqref{eq:G13G31_pair_app} into \eqref{eq:Ctilde_def_app}, and omitting the $G_{33}^*$ contribution and the higher-order far-field remainders, the projected pair mobility can be written, at the order retained here, as
\(\widetilde{\mathcal C}^*(\bm r_1^*,\bm r_2^*)=\mathcal C^*(t^*)+O((r_\parallel^*)^{-5})\), where
\begin{equation}
\mathcal C^*(t^*)
\coloneqq
\frac{3}{2\pi}z_1^*z_2^*
\left[
\frac{\cos^2\phi}{(\ell^*+\xi_-^*)^3}
+
\frac{\sin^2\phi\,\xi_-^*}{(\ell^*+\xi_-^*)^4}
\right].
\end{equation}

\section{Evaluation of the first-order mean flux} \label{app:first_order_flow}
In this appendix, we summarize the analytical evaluation of the period average in \eqref{eq:Q1_by_parts}.
The calculation is straightforward but lengthy, and is therefore collected here.
We begin from the leading-order solution \eqref{eq:xi0_sol} and introduce
\begin{equation}
A_c^* \coloneqq A^*\cos\!\left(\frac{\delta}{2}\right),
\qquad
A_s^* \coloneqq A^*\sin\!\left(\frac{\delta}{2}\right),
\qquad
\vartheta^* \coloneqq \chi^*-\frac{\pi}{2}
\label{eq:Acbtheta_def}
\end{equation}
and, we have
\begin{equation}
\xi_{+,0}^*=-A_c^* \sin\vartheta^*,
\qquad
\xi_{-,0}^*=A_s^*\cos\vartheta^*.
\label{eq:xi0_theta}
\end{equation}
From \eqref{eq:W0_def}, one obtains
\begin{equation}
W_0^*
=
-A_c^*\sin\vartheta^*
-\frac{\tan\phi}{4}
\left(
A_c^{*2}\sin^2\vartheta^*-A_s^{*2}\cos^2\vartheta^*
\right),
\label{eq:W0_theta}
\end{equation}
and hence
\begin{equation}
\dot W_0^*
=
-\omega^*
\left[
A_c^*\cos\vartheta^*
+\frac{\tan\phi}{4}A^{*2}\sin{2\vartheta^*}
\right].
\label{eq:dW0_theta}
\end{equation}
Likewise,
\begin{equation}
z_1^*z_2^*
=
1+\tan\phi\,A_c^*\sin\vartheta^*
+\frac{\tan^2\phi}{4}
\left(
A_c^{*2}\sin^2\vartheta^*-A_s^{*2}\cos^2\vartheta^*
\right),
\label{eq:zprod_theta}
\end{equation}
and
\begin{equation}
z_2^*-z_1^*
=
-\tan\phi\,A_s^*\cos\vartheta^*.
\label{eq:zdiff_theta}
\end{equation}
Substituting \eqref{eq:zprod_theta} and \eqref{eq:zdiff_theta} into
\eqref{eq:Cstar_app}, the coupling coefficient becomes
\begin{equation}
\mathcal C^*(t^*)
=
\frac{3}{2\pi}\,
z_1^*z_2^*
\left[
\frac{\cos^2\phi}{(\ell^*+A_s^*\cos\vartheta^*)^3}
+
\frac{\sin^2\phi\,A_s^*\cos\vartheta^*}{(\ell^*+A_s^*\cos\vartheta^*)^4}
\right].
\label{eq:C_theta}
\end{equation}
Substituting \eqref{eq:dW0_theta} and \eqref{eq:C_theta} into \eqref{eq:Q1_by_parts}, the integrand becomes a rational trigonometric function of \(\vartheta^*\).
Since the denominator depends only on \(\cos\vartheta^*\), all terms odd in \(\sin\vartheta^*\)
vanish upon integration over one period.
It is therefore convenient to write
\begin{equation}
\begin{aligned}
z_1^*z_2^*
&=
c_0+c_1\sin\vartheta^*+c_2\cos 2\vartheta^*,
\\
z_1^*z_2^*(z_2^*-z_1^*)
&=
c_3\cos\vartheta^*+c_4\sin 2\vartheta^*+c_5\cos 3\vartheta^*,
\\
\frac{1}{\omega^*}\dot W_0^*
&=
c_6\cos\vartheta^*+c_7\sin 2\vartheta^*,
\end{aligned}
\label{eq:PMR_harmonics}
\end{equation}
where
\begin{equation}
\begin{aligned}
c_0
&=
1+\frac{\tan^2\phi}{8}\left(A_c^{*2}-A_s^{*2}\right),
&
c_1
&=
\tan\phi\,A_c^*,
&
c_2
&=
-\frac{\tan^2\phi}{8}A^{*2},
\\
c_3
&=
-\tan\phi\,A_s^*
\left[
1+\frac{\tan^2\phi}{16}\left(A_c^{*2}-3A_s^{*2}\right)
\right],
&
c_4
&=
-\frac{\tan^2\phi}{2}A_c^*A_s^*,
&
c_5
&=
\frac{\tan^3\phi}{16}A^{*2}A_s^*,
\\
c_6
&=
-A_c^*,
&
c_7
&=
-\frac{\tan\phi}{4}A^{*2}.
\end{aligned}
\label{eq:PMR_coeffs}
\end{equation}
Reducing the remaining trigonometric polynomials to harmonics in \(\cos(m\vartheta^*)\), we introduce the integrals
\begin{equation}
K_m^{(p)}(\ell^*,A_s^*)
\coloneqq
\int_0^{2\pi}
\frac{\cos(m\vartheta^*)}{(\ell^*+A_s^*\cos\vartheta^*)^p}\,
\mathrm d\vartheta^*,
\qquad
|A_s^*|<\ell^*.
\label{eq:Kmp_def}
\end{equation}
The condition \(|A_s^*|<\ell^*\) ensures that the denominator does not vanish over one period.
With these definitions, the period average in \eqref{eq:Q1_by_parts} becomes
\begin{equation}
\begin{split}
Q_1^*
=
-\frac{3\omega^*\cos^2\phi}{8\pi^3(\gamma_0^*)^2}
\Bigg[
&
\Bigl(2c_0c_6+c_1c_7+c_2c_6\Bigr)K_1^{(3)}
+
\Bigl(c_2c_6-c_1c_7\Bigr)K_3^{(3)}
\\
&\hspace{1.6cm}
-\tan\phi\Bigl(c_3c_6+c_4c_7\Bigr)K_0^{(4)}
-\tan\phi c_6\Bigl(c_3+c_5\Bigr)K_2^{(4)}
\\
&\hspace{1.6cm}
-\tan\phi\Bigl(c_5c_6-c_4c_7\Bigr)K_4^{(4)}
\Bigg].
\end{split}
\label{eq:Q1_reduced_preK}
\end{equation}
At the order retained in the main text, the final term in \eqref{eq:Q1_reduced_preK}, involving \(K_4^{(4)}\), is beyond the asymptotic accuracy considered here and will be neglected hereafter.
The remaining required integrals can be generated from the standard formula
\begin{equation}
I_m(\ell^*,A_s^*)
\coloneqq
\int_0^{2\pi}
\frac{\cos(m\vartheta^*)}{\ell^*+A_s^*\cos\vartheta^*}\,
\mathrm d\vartheta^* =
\frac{2\pi}{\sqrt{\ell^{*2}-A_s^{*2}}}
\left(
\frac{\sqrt{\ell^{*2}-A_s^{*2}}-\ell^*}{A_s^*}
\right)^m,
\qquad
|A_s^*|<\ell^*,
\label{eq:Im_def}
\end{equation}
which may be found, for example, in \citet{gradshteyn2014table}.
For \(p\ge2\), differentiating \eqref{eq:Im_def} with respect to \(\ell^*\) gives
\begin{equation}
K_m^{(p)}(\ell^*,A_s^*)=
\frac{(-1)^{p-1}}{(p-1)!}
\frac{\partial^{p-1}}{\partial \ell^{*\,p-1}}
I_m(\ell^*,A_s^*).
\label{eq:Kmp_from_Im}
\end{equation}
For the present calculation, the required exact forms are
\begin{equation}
\begin{aligned}
K_1^{(3)}
&=
-\frac{3\pi A_s^*\ell^*}{(\ell^{*2}-A_s^{*2})^{5/2}},
\\
K_3^{(3)}
&=
\frac{\pi}{A_s^{*3}}
\left[
8+
\frac{-15A_s^{*4}\ell^*+20A_s^{*2}\ell^{*3}-8\ell^{*5}}
{(\ell^{*2}-A_s^{*2})^{5/2}}
\right],
\\
K_0^{(4)}
&=
\frac{\pi \ell^*(2\ell^{*2}+3A_s^{*2})}{(\ell^{*2}-A_s^{*2})^{7/2}},
\\
K_2^{(4)}
&=
\frac{5\pi A_s^{*2}\ell^*}{(\ell^{*2}-A_s^{*2})^{7/2}}.
\end{aligned}
\label{eq:K_exact_all}
\end{equation}
Expanding \eqref{eq:K_exact_all} for \(A_s^*/\ell^*\ll1\), and
substituting the resulting series into \eqref{eq:Q1_reduced_preK} after omitting the
\(K_4^{(4)}\) term, one directly obtains the first-order mean flux \eqref{eq:Q1_final}--\eqref{eq:H2_def} in the main text.

\section{Coordination under a background shear flow}
\label{app:shear_phase_selection}

\begin{figure}
    \centering
    \includegraphics[width=0.8\linewidth]{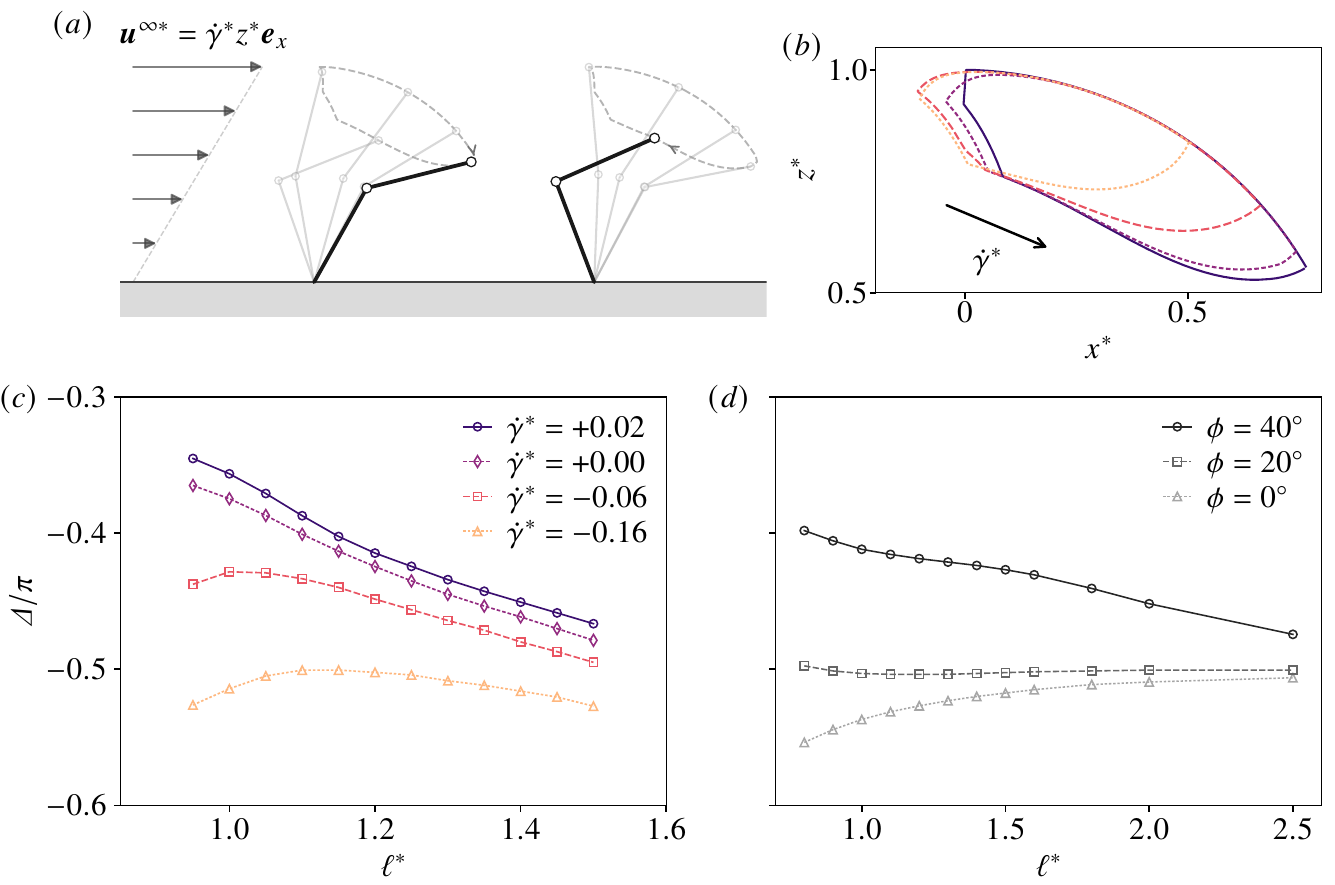}
\caption{
Flow-maximising coordination under external modification of the beating orbit by a background simple shear flow.
(a) Schematic of the two cilia system under a background flow with a dimensionless shear rate $\dot{\gamma}^*$.
(b) Tip-bead trajectories of a single cilium, optimised by reinforcement learning under different shear rates. The bending stiffness is set to \(k_b^*=1.0\).
(c,d) Flow-maximising phase difference \(\Delta/\pi\) as a function of the inter-cilium spacing \(\ell^*\) for (c) the two cilia system, and (d) the tilted-slider model.
The conditions are \(k_b^*=1.0\) for the two cilia, and \(a^*=0.05\), \(k_r^*=2\) and \(\omega^*=\pi\) for the slider model. 
\label{fig:shear_phase_sweep}
}
\end{figure}

The tilted-sliders in figure~\ref{fig:fig6} show that the flow-maximising phase difference depends not only on the inter-cilium spacing but also on the effective beat geometry.
As a complementary check in the cilium model, we briefly examine how the optimal coordination is altered by externally modifying the beating orbit with a background simple shear flow \(\boldsymbol{u}^{\infty}=\dot{\gamma} z\,\boldsymbol{e}_x\), where $\dot{\gamma}$ is the shear rate as shown in figure~\ref{fig:shear_phase_sweep}(a).
Using the cilium length scale \(L\) and the velocity scale \(\tau_{\max}/(\mu L)\), the shear flow can be described in a dimensionless form as
\(\boldsymbol{u}^{\infty *}=\dot{\gamma}^* z^*\boldsymbol{e}_x\), with \(\dot{\gamma}^*=\mu L^2\dot{\gamma}/\tau_{\max}\).
Note that the imposed shear flow is introduced solely to perturb the realised beat geometry, rather than to model ciliary pumping in an externally sheared environment.

For each value of \(\dot{\gamma}^*\), an isolated cilium was first optimised by reinforcement learning under the imposed shear flow, and the resulting tip-bead trajectories are shown in figure~\ref{fig:shear_phase_sweep}(b).
As the shear rate \(\dot{\gamma}^*\) decreases, the orbit is lifted away from the wall and becomes closer to a wall-parallel stroke, whereas the trajectory at \(\dot{\gamma}^*=0.02\) is more strongly tilted.
The torque sequence obtained with a single cilium was then used in the two cilia system under the same shear flow \(\dot{\gamma}^*\), and the flow-maximising coordination is evaluated using the phase-sweep analysis described in \S\ref{subsec:twocilia}.
Figure~\ref{fig:shear_phase_sweep}(c) shows that the applied shear flow changes the shape of the $\ell^*$--$\Delta$ relation: when the shear rate is large, the flow-maximising phase difference $|\Delta|$ monotonically grows with the inter-cilium spacing $\ell^*$, whereas for small shear rates, the trend becomes non-monotonic with a change in slope around $\ell^* \sim 1.1$.
In other words, for short inter-cilium spacings $\ell^* \lesssim 1.1$, the sign of the slope in the $\ell^*$--$\Delta$ relation depends on the beating orientation: tilted beating orbits give a negative slope, whereas nearly wall-parallel strokes give a positive slope.
The same trend can also be found in the tilted-slider model shown in figure~\ref{fig:shear_phase_sweep}(d), and the slope in the $\ell^*$--$\Delta$ relation depends on the beating angle $\phi$.
Since the tilted-slider model captures the qualitative trend in the flow-maximising phase difference under variations in the beating pattern, these results suggest that such reduced models can help rationalise how flow-maximising coordination depends on the beating pattern.

\bibliography{reference}

\end{document}